\documentclass[fleqn,10pt]{wlscirep}
\usepackage[utf8]{inputenc}
\usepackage[T1]{fontenc}
\usepackage{subcaption}
\usepackage{graphicx}
\usepackage{booktabs}
\usepackage{amssymb}
\usepackage{subcaption}
\usepackage{caption}
\usepackage{float}
\usepackage{placeins}
\usepackage{xcolor}
\usepackage{ulem}
\usepackage{amsmath}

\usepackage{pdfpages}

\title{Probabilistic neighbors' selection competes with confirmation bias in a bounded confidence opinion model}
\author[1,2,3,*]{Chiara Giaquinta}
\author[2]{Laura Hernández}
\author[1,3]{David Chavalarias}

\affil[1]{CNRS, Complex Systems Institute of Paris Île-de-France (ISC-PIF) Paris, France}
\affil[2]{Laboratoire de Physique Théorique et Modélisation, CNRS-CY Cergy Paris Université, Cergy-
Pontoise, France}
\affil[3]{EHESS, Centre d’Analyse et de Mathématique Sociales (CAMS) Paris, France}

\affil[*]{chiara.giaquinta@cyu.fr}

\begin{abstract}
In this work, we investigate three modified versions of the classic Hegselmann-Krause opinion dynamics model, incorporating features that are typical of many real-world systems, such as uncertainty in the selection of interacting agents and a weighted evaluation of the relevance of their opinions in the influence function to enhance confirmation bias. Through extensive simulations across different network topologies, ranging from stylized network models—Barabási-Albert, Erdős-Rényi, and Watts-Strogatz networks—to empirical networks with community structure, we identify the influence of each modification on opinion evolution and convergence. Our findings reveal that they exert opposite effects on the bounded confidence threshold required for consensus. We further explore an extension based on a data-driven opinion initialization on the empirical networks, where initial opinions are drawn from Gaussian distributions specific to each detected community. While the qualitative effects of the three modified models remain consistent, this new initialization strategy reveals distinct dynamics within the network communities. These insights provide a new and comprehensive perspective on how realistic variations of the Hegselmann-Krause model, in terms of both interaction rules and initial opinion distributions, affect opinion dynamics and shed light on the mechanisms of consensus formation within structured communities.
\end{abstract}
\begin{document}

\flushbottom
\maketitle
\thispagestyle{empty}

%\noindent Please note: Abbreviations should be introduced at the first mention in the main text – no abbreviations lists. Suggested structure of main text (not enforced) is provided below.

\section{Introduction}\label{sec:introduction}

Among the different mathematical models devoted to the study of the formation and diffusion of opinions in a society~\cite{castellano_statistical_2009}, \textit{bounded confidence} models \cite{lorenz_continuous_2007} account for two important ingredients of social interactions: \textit{homophily} and \textit{social influence}. These models posit that social agents only choose to interact with others whose opinions are not significantly different from their own, disregarding those holding very distinct viewpoints. As a consequence of these interactions, the opinions of the interacting agents become increasingly similar. The two most popular bounded confidence models address these processes in different ways: the Deffuant-Weisbuch (DW) model, which is based on sequential pairwise interactions \cite{deffuant_mixing_2000}, and the Hegselmann-Krause (HK) model, in which all agents update their opinions synchronously by taking the average of their own opinion and those of their neighbors \cite{hegselmann_opinion_2002}. 

In the HK model \cite{hegselmann_opinion_2002}, the choice of the neighbors that will interact with each agent is strictly conditioned by a model parameter, $\varepsilon$, called \textit{confidence bound}: only agents whose opinion differs from that of the active agent by less than $\varepsilon$ are included in its set of interacting neighbors. This mechanism provides a simplified yet effective representation of certain real-world scenarios, such as the diffusion of information on an online platform, where agents scroll through their timeline and receive information and influence from a limited set of peers who are either proposed by the algorithm, selected by the agent, or a mixture of both.

While the original HK model provides a valuable framework for studying opinion dynamics, numerous extensions have been proposed to relax some of its simplifying assumptions or to investigate its mathematical properties. Several extensions of the HK framework have been developed to account for more realistic social behaviors. One line of research introduced heterogeneous confidence bounds, allowing agents to differ in their openness to alternative opinions, and showed that such heterogeneity may promote consensus and shift it away from the center of the opinion space ~\cite{fu_opinion_2015,lorenz_heterogeneous_2010}. A different line of work replaced the binary interaction rule with smooth influence functions and investigated the role of extremist agents in the emergence of opinion shifts toward extreme positions \cite{deffuant_modelling_2004}. Continuous-time formulations have also been introduced to establish rigorous convergence results for finite-agent and continuum versions of the HK dynamics \cite{blondel_v_d_continuous-time_2010}. 

Recent studies have actively sought to incorporate widely observed characteristics of real-world social systems. These modifications aim to bridge the gap between theoretical studies of opinion dynamics and empirical observations, encompassing aspects such as networked societies \cite{fortunato_consensus_2005,schawe_when_2021}, heterogeneous agents \cite{schawe_when_2020, perrier_phase_2024}, stubborn agents \cite{ghaderi_opinion_2014}, agents that bear a cost in changing their opinions~\cite{schawe_collective_2020} and mass media effects \cite{pineda_mass_2015}, to cite a few.

The deterministic selection of interacting neighbors in the original HK model warrants examination, as it leads to dynamics that are strongly affected by the hard cutoff in neighbor selection~\cite{lorenz_continuous_2007}. Conceptually, the model assumes that: ($A_1$) stylized opinions can be represented by real numbers in the $[0,1]$ interval, ($A_2$) social actors can be modeled as entities that exhibit threshold behavior, and ($A_3$) they can infer the opinions of others without error, so that each one can decide with whom to interact. 

While $A_1$ is a simplified way to represent a continuum of opinions between two extremes and is common to various opinion dynamics models, several recent works have pointed out the discrepancy between the opinions actually held by agents and those expressed by them, which challenges both $A_2$ and $A_3$, leading to the development of models with varying degrees of complexity that incorporate this ambiguity \cite{dong_opinion_2024,jarema_private_2022,hou_opinion_2021,gaisbauer_dynamics_2020}. A different approach to address this issue is to introduce noise. Noisy bounded confidence models have been studied, for instance, by incorporating both $\textit{selectivity}$ and $\textit{adaptation}$ noises through random fluctuations added, respectively, to the confidence bound $\varepsilon$ and to the updated opinion \cite{steiglechner_noise_2024}. Similarly, other studies have introduced random noise exclusively in the opinion update process \cite{goddard_noisy_2022, mas_individualization_2010}. For the Deffuant-Weinsbuch model, a model with a fuzzy selection of interacting pairs of agents, along with a random, spontaneous change in agents' opinions, was studied~\cite{grauwin_opinion_2012,deffuant_modelling_2004}.

Interestingly, empirical evidence also challenges assumptions $A_2$ and  $A_3$. Chavalarias et al. \cite{chavalarias_can_2024} studied the retweet patterns extracted from Twitter during the 2017 and 2022 presidential elections in France. Considering these patterns as a proxy for inter-individual influence and opinion alignment, they found that the probability of interaction is a decreasing function of the opinion distance between two agents rather than a binary threshold function.

In order to account for this aspect of real systems, we replace the sharp threshold in the HK model with a \textit{probability of selection of interacting agents}, parametrized by the confidence bound, $\varepsilon$. In this way, there is a non-zero probability that the active agent could be influenced by others who are further away in opinion space. On the other hand, there is also a non-zero probability of rejecting an interaction with agents whose opinion difference is close to the  $\varepsilon$ threshold.

Another aspect of the classic HK model that we want to question here is the fact that, in the original model, all the neighbors interacting with a given agent exert the same influence on it during the opinion update, regardless of their opinion difference. It is clear that this is also an unrealistic assumption, whose effects need to be studied. Within the bounded confidence framework, which can be interpreted as a stylized representation of \textit{confirmation bias}~\cite{lord_biased_1979}, it is reasonable to expect that agents whose opinions are closer to that of the active agent will influence it more than those whose opinions are much further away. Early studies within the bounded confidence framework proposed replacing the binary interaction rule with smooth influence functions, typically by weighting neighboring opinions through a Gaussian function of their opinion distance, with a variance related to the confidence bound~\cite{deffuant_modelling_2004}. More recently, generalized bounded confidence models have also adopted smooth weighted influence functions in continuous-time formulations, enabling analytical investigations of their stability properties~\cite{brooks_emergence_2023}. Most of these studies focused on the DW model and considered either continuous-time formulations, mean-field settings, or simulations on relatively small and highly structured networks. 

Specifically, in this work we systematically study the effect of fuzziness in the selection of interacting neighbors, as well as that of a weighted dynamical influence function that depends on the opinion difference and modulates the relative influence that the different selected agents have during the discrete-time opinion update process of the HK model. We compare models that consider each of these aspects separately, as well as their combination. 
 
We also investigate the role of the topology of the potential interactions among agents, by studying all the considered models on two types of networks. On the one hand, we consider synthetic networks with well-controlled properties, such as homogeneous or heterogeneous degree distributions, clustering, and small-world effecs (Barabási-Albert, Erdős-Rényi, and Watts-Strogatz networks). On the other hand, we consider empirical networks representing online social interactions obtained from a massive data collection of Twitter data carried out while access to the platform was open for research \cite{noauthor_projet_nodate}.

The paper is structured as follows: in Section \ref{sec:models and methods}, we provide a detailed description of the three modified HK models, the properties of both the synthetic and the empirical networks studied, and the implementation of different opinion initialization methods used in this work. Section \ref{sec:results} presents our main findings, including a comparative analysis of the steady states of the dynamics of each of the models for different values of the confidence bound, network topology, and opinion initialization. In Section \ref{sec:discussions} we discuss how our main findings compare with other variants of the model and propose some directions for further research. Finally, Section \ref{sec:conclusions} presents a summary of the results and the conclusions of this study.

\section{Methods}\label{sec:models and methods}

We present a systematic study of three variants of the original HK model (hereafter denoted $HK_0$). In two of them, we study the effect of a single modification of the dynamical rules (weighted influence and fuzzy neighbor selection, respectively), while in the third we study the result of combining both modifications.

\subsection{Dynamical rules for the three studied models}\label{sec:models and methods,model dynamics}

The main modifications introduced to the dynamical rules concern the selection of the interacting neighbors and their influence on the active agent.

\paragraph*{Selection of the interacting  neighbors. }

The potential interacting neighbors are given by the different types of networks we consider. The selection of the neighbors that will actually interact with each active agent $i$ can occur in two possible ways:

\begin{itemize}
    \item $\textbf{Strict Selection}$: Agent $i$ interacts with a topological neighbor $j$ at time $t$ if their opinion difference is below the confidence threshold: $|x_i(t) - x_j(t)| < \varepsilon$. This selection criterion corresponds to the strict step function condition of the $HK_0$ model.
    \item $\textbf{Fuzzy Selection}$: Agent $i$ interacts with a neighbor $j$ at time $t$ with a probability that depends on their opinion difference, $f(x_{ij}(t))$, given by Equation~\ref{eq:smooth_function}. 
    
          \begin{equation}
\centering
    f(x) = \frac{1}{1 + e^{\frac{\beta}{\varepsilon}(x - \varepsilon)}} \label{eq:smooth_function}
\end{equation}

where $x = x_{ij}(t) = |x_i(t)-x_j(t)|$ corresponds to the absolute value of the opinion difference between a generic agent $i$ and its neighbor $j$ at time $t$. $\varepsilon \in [0,1]$ is the $\textit{confidence interval}$ of the $HK_0$ model, and $\beta$ is a parameter controlling the steepness of the slope in the region of strong variation of $f(x)$. Higher values of $\beta$ push the function closer to the sharp step function characteristic of the $HK_0$ model, while lower values produce the desired softening of the curve. Fig.~\ref{smooth fct} illustrates our choices of $\beta$ values.  
    
More specifically, the interaction between $i$ and $j$ occurs only if $r < f(x_{ij}(t))$, where $r$ is a random number uniformly drawn from the interval $[0,f(x_{ii}(t))]$. This new selection criterion accounts for the intrinsic uncertainty in the perception of opinion differences.

\begin{figure}
            \centering
            \includegraphics[width=0.5\textwidth]{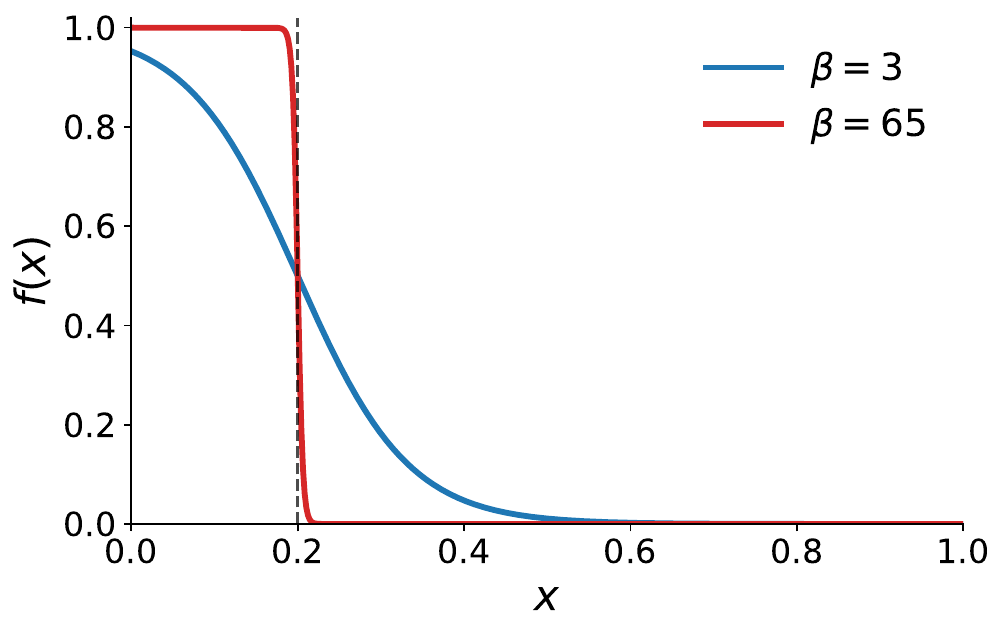}
            \caption{For $\varepsilon=0.2$, the $f(x)$ of Eq.~\ref{eq:smooth_function} for two different $\beta$ values. Higher values of $\beta$ push the function closer to the sharp step function characteristic of the $HK_0$ model, while lower values produce the desired softening of the curve. The vertical dotted line denotes the $\varepsilon=0.2$ value.}
            \label{smooth fct}
            \end{figure}

\end{itemize}

\paragraph*{Weighting the influence of the selected neighbors}
Let $I(t)$ be the set of neighbors selected by agent $i$ at time $t$, as defined above. Note that agent $i$ is also included in $I(t)$, meaning that agent $i$ considers itself its own neighbor. The active agent $i$ can then update its opinion in two ways, according to the model considered:

\begin{itemize}
    \item $\textbf{Simple Mean Average}$: Agent $i$ assigns the same weight to each interacting neighbor, performing an arithmetic mean, as in the original $HK_0$ model:
    \begin{equation}
        x_i(t+1) = \frac{\sum_{j \in I(t)} x_j(t)}{|I(t)|}
    \end{equation}
    where $|I(t)|$ is the cardinality of $I(t)$.
    \item $\textbf{Weighted Average}$: Agent $i$ assigns different weights to the opinions of its interacting neighbors, using Eq.~\eqref{eq:smooth_function} to perform a weighted average:
       \begin{equation}
        x_i(t+1) = \frac{\sum_{j \in I(t)} x_j(t)f(x_{ij}(t))}{\sum_{j \in I(t)} f(x_{ij}(t))}
    \end{equation}
\end{itemize}

In this  setting, the weight agent $i$ assigns to itself is $f(x_{ii}(t)) = \frac{1}{1+e^{-\beta}}$, $\forall t$.

\paragraph{Description of the three studied models}

\begin{itemize}
    \item \textit{Model} $HK_1$: To address the issue that in the $HK_0$, all the selected neighbors have the same influence on the active agent, regardless of how far their opinions are, we weight the influence that each of them has on the active agent using Eq.~\ref{eq:smooth_function}, thus enhancing confirmation bias. The selection of the interacting neighbors instead is performed strictly, as in the $HK_0$ model. 
    \item \textbf{Model} $HK_2$: We relax the hypothesis of a strict threshold for the selection of the interacting neighbors, changing it into a fuzzy selection, as was done for small systems with pairwise, sequential interactions in Ref.~\cite{grauwin_opinion_2012}. This variant introduces uncertainty in the perception of opinion differences by means of a finite probability that agents slightly outside the confidence interval might be included, while some agents inside, but near the confidence threshold, might be ignored in the interaction. However, once the interacting agents have been chosen, their influence on the active agent is the same for all of them. 
    \item \textbf{Model} $HK_3$: This model combines both the fuzzy selection of the interacting neighbors and their weighted influence.
  
\end{itemize}

 The different studied models are summarized in Table~\ref{tab:overview models}.

\begin{table}[h]
    \centering
    \caption{Models overview}
    \label{tab:overview models}
    \begin{tabular}{ccccc}
        \toprule
         & $HK_0$ & $HK_1$ & $HK_2$ & $HK_3$ \\
        \midrule
        $\textbf{Selection}$ & Strict & Strict & Fuzzy & Fuzzy \\
         &  &  &  &  \\
        $\textbf{Opinion Update}$ & Mean & Weighted & Mean & Weighted \\
                       & average & average & average & average \\
        \bottomrule
    \end{tabular}
\end{table}

For each model, the opinion initialization is performed according to the methods detailed in Section \ref{sec:opinion initialization}. 

At each time step $t$, all $N$ agents update their opinions synchronously. The generic active agent $i$ first selects the interacting agents among its topological neighbors, according to the rule corresponding to each model, and then it decides if and how to weight each of the interacting neighbor's opinions. 

The simulation stops when there are no significant changes in the agents' opinions between two consecutive time steps, satisfying the following convergence criterion. 

\begin{equation}
    \sum_{i=1}^{N} |x_{i}(t-1) - x_{i}(t)| < 10^{-2} \label{criterion convergence}
\end{equation}

It should be noted that in the case of fuzzy selection of interacting neighbors (models $HK_2$ and $HK_3$), this criterion does not guarantee full convergence, because there is a non-zero probability that, if we continue the simulation for a sufficiently long time, the set of interacting agents might at some stage include an agent whose opinion is very far from that of the active agent, which could re-trigger the dynamics. This scenario, although possible, is very unlikely and was not observed within the maximum running times of our simulations. Moreover, tests with a stricter convergence criterion do not show qualitatively different results (for further details, see Section S1 of the Supplementary Material).

\subsection{Networks under study}\label{sec:models and methods,studied networks}

All models are implemented on a static graph $G=(V,E)$, where $V$ represents the set of $N$ vertices (agents) and $E$ the set of undirected edges (social links between agents). 

We analyzed the behavior of our models across several types of networks, to understand the impact of structural differences on opinion dynamics, if any. Specifically, we studied three stylized network models—Barabási-Albert (BA), Erdős-Rényi (ER), and Watts-Strogatz (WS)— along with two empirical networks known to have a clear community structure and collected from Twitter data before access to the platform became restricted for researchers. These networks, labeled as GI and GP, are built from retweet data corresponding to two different discussion subjects, one concerning "immigration and multiculturalism" and the other "foreign politics", that took place during the year 2017 in France. They are part of the data analyzed in the \textit{Politoscope project}, collected using the MULTIVAC platform ~\cite{noauthor_projet_nodate}. The empirical retweet networks were originally weighted and directed. After the preprocessing steps described in Section S2 of the Supplementary Material, they were converted into unweighted and undirected graphs. Although retweet interactions are inherently directed, we use the resulting undirected networks as proxies for the underlying social interaction structure through which opinions may spread. Since our goal is to study opinion dynamics rather than information diffusion on Twitter itself, we consider this approximation appropriate. For concision, we mainly present the results corresponding to the GI network, but similar results are found for the GP network as well. For further details on the construction of both empirical networks, as well as the community detection algorithm used on them, refer to Section S2 of the Supplementary Material.

To ensure meaningful comparisons with the real-world network GI (characterized by a size $N=53002$ and average degree $<k>=6.3$), the stylized networks were generated with similar parameters: a size of $N=50000$ vertices and an average degree $<k> \approx 6$. Unlike the artificial networks, the empirical network features strong community structures that, as shown in Section \ref{sec:results}, significantly influence the outcomes of the dynamics.

Final opinion clusters are composed of agents holding the same opinion within a tolerance range of $10^{-2}$. As usual, we define the order parameter as the average normalized size of the largest opinion cluster, $<S>$, which, unless stated otherwise, is taken over $100$ independent realizations. Each realization is characterized by a different network randomization (except for the empirical case) and a different set of initial opinions.

According to the value of the order parameter $<S>$, the system is categorized into the following states:

\begin{itemize}
    \item $\textit{Unanimity}$, if $<S>=1$. The entire population holds the same final opinion within the defined precision.
    \item $\textit{Consensus}$, if $<S> \lessapprox 1$. A single, dominant opinion cluster encompasses nearly the entire population.
    \item $\textit{Polarization}$, if $<S> \approx 0.5$, coexisting with another cluster of similar size.
    \item $\textit{Fragmentation}$, if $<S> \ll 1$. Many small, distinct clusters are formed, and no major opinion cluster dominates.
\end{itemize}

\subsection{Opinion initialization and other parameters}\label{sec:opinion initialization}

Following a standard practice in ABM simulations of opinion dynamics models, we use uniformly distributed opinions as initial conditions, $x_i(t=0) \in [0,1]$, for the three stylized network models studied here (BA, ER, and WS). However, for the study of the models on empirical networks, it is necessary to consider the origin of their structure. Unlike in theoretical models, where the topology of the interactions is imposed by a fixed network and then the initially random distributed opinions evolve through repeated interactions, online social networks have actually been formed through the co-evolution of the agents' opinions and their connections. In other words, the network is shaped by homophily and displays a strong community structure. In this situation, a uniform distribution of initial opinions in a fixed network may seem unrealistic. Therefore, along with the standard uniform distribution of initial opinions, we also explore the case where agents within the same community are, on average, closer in opinion space, than those belonging to different ones. To do so, we use the detected community structure and draw the initial opinions of agents belonging to the same community from a Gaussian distribution. These distributions have different means, and we use our dataset to parametrize some of them. The French political landscape in 2017 was dominated by three political tendencies: center (C), extreme right (XR) and left (L), corresponding to the three candidates who obtained the largest number of votes respectively. We identify the communities associated to these three political tendencies, and initialize opinions for each of them in the following way:
 \begin{itemize}
     \item We set the mean opinion of each political community $c$ as $\mu_{c=L}=0.2$, $\mu_{c=C}=0.5$, $\mu_{c=XR}=0.8$.
     \item We initialize the opinions of the agents in each political community with a Gaussian distribution centered on the corresponding mean: $\mathcal{N}(\mu_{c}, 0.05^2)$. Since opinions are restricted to the interval $[0,1]$, sampled values lying outside this range are clipped to the corresponding boundary.
     \item For all the other communities, their mean $\mu_c$ is sampled randomly from a uniform distribution in the interval $[0.1, 0.9]$, and agents in each community $c$ draw their initial opinions from $\mathcal{N}(\mu_c, 0.05^2)$. Importantly, these means are redefined for each independent realization, while those corresponding to the communities of the three main political leaders remain constant.
 \end{itemize}

\section{Results}\label{sec:results}

We simulate the different dynamics indicated in Table~\ref{tab:overview models} for several independent realizations of each network type, as described in Section \ref{sec:models and methods}. These realizations are obtained by randomizing both the initial opinions and the network structure for the theoretical networks, and only the initial opinions for the empirical networks. More precisely, for the latter, taking into account the role of homophily in real systems, which shapes the networks we observe, we have tested two types of random initial conditions: uniform distribution and a truncated Gaussian distribution, centered around the opinions of the leaders.

For this setup, and unless stated otherwise, we performed $100$ independent realizations for each model and network topology, running each simulation with the convergence criterion given by Eq.~\ref{criterion convergence} and with a maximum of $10^5$ time steps, except for the Gaussian initial conditions where, due to the long convergence times observed with this structured initialization, the maximum number of steps was increased to $3 \cdot 10^5$, thus ensuring that all realizations had satisfied the criterion in Eq.~\ref{criterion convergence}.

The parameter space is explored in the region of $\varepsilon \in [0.02, 0.33]$, where the transition to consensus is expected, with a step size of $\Delta \varepsilon = 0.01$,  while keeping the steepness parameter $\beta=3$ for all simulations. The specific choice of $\beta=3$ corresponds to a moderate "softening" of the step function of the $HK_0$ model (as shown in Figure \ref{smooth fct}).

We start by discussing the ensemble of results corresponding to simulations with uniformly distributed initial opinions, both in the stylized and empirical networks.

\begin{figure*}
    \centering
        \includegraphics[width=\textwidth]{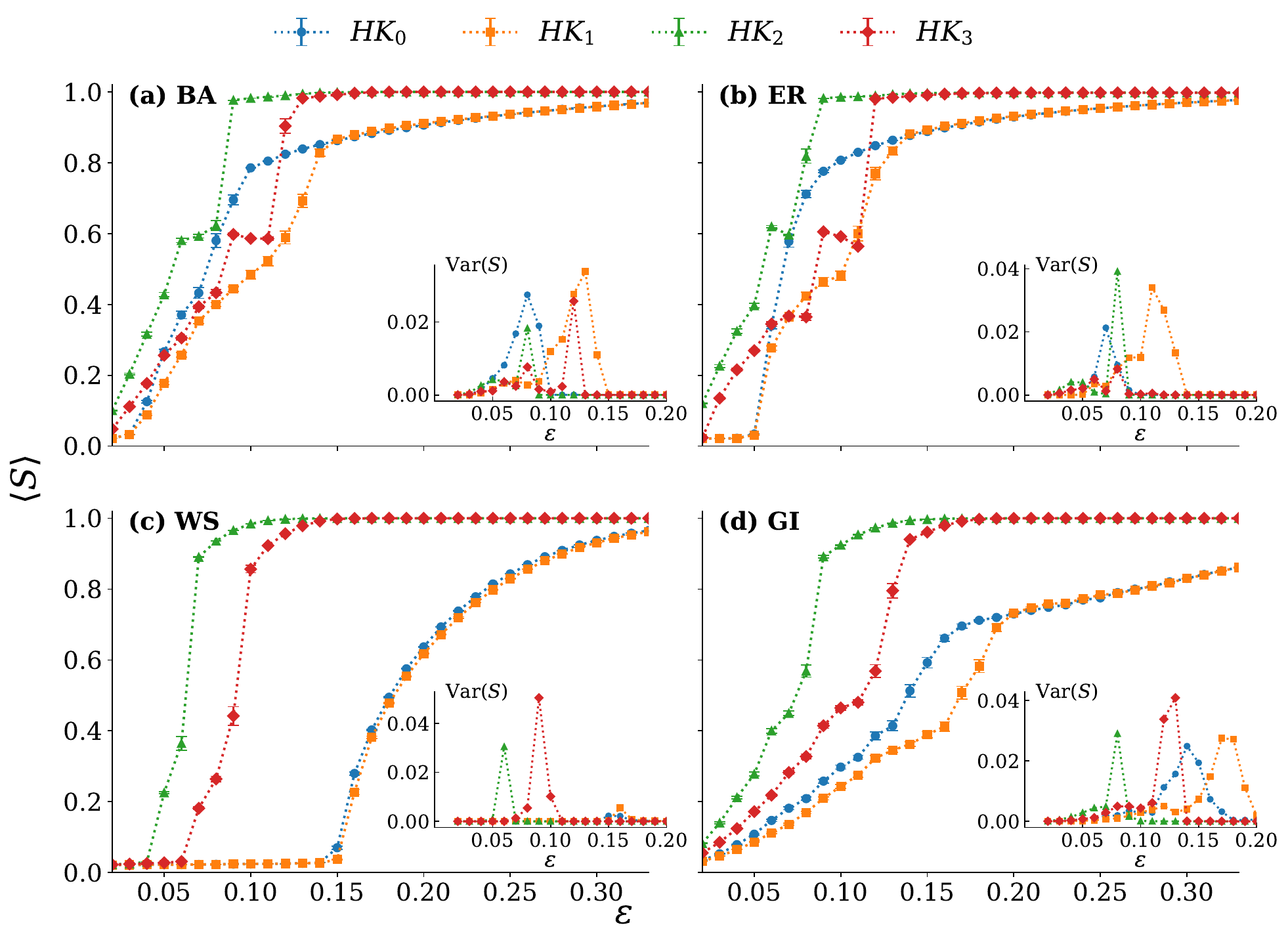}
        \caption{Order parameter $<S>$ as a function of $\varepsilon$ for $\beta=3$ and uniform initial conditions, for different network topologies (BA, ER, WS and GI). Curves correspond to different models, $HK_0$–$HK_3$. Insets display the variance of $<S>$, used to identify the critical threshold $\varepsilon_c$. The introduction of the probabilistic selection of the interacting neighbors induces a sharp transition to consensus at relatively low $\varepsilon$ values, while the enhanced confirmation bias has the effect of hindering consensus with respect to the original $HK_0$. The transition is sharp in random and empirical networks but becomes a smooth crossover in the WS network. Dotted lines are only a guideline for the eye.}
    \label{picture 1}
\end{figure*}

Fig.~\ref{picture 1} summarizes the main results of this part of the work. The insets in the different panels show the variance of the order parameter, whose peak allows us to determine the particular value of the confidence interval that signals the transition to consensus and that is usually called \textit{critical} confidence interval, $\varepsilon_c$. 

The comparison of the different models for a fixed topology reveals the two major effects of the introduction of the probability of selection of the interacting neighbors ($HK_2$) and of the enhanced confirmation bias in the influence function ($HK_1$): \textit{the former induces a sharp transition to consensus at relatively low values of $\varepsilon$, while the latter hinders consensus, increasing the value at which a large opinion cluster emerges, with respect to the original $HK_0$.} With the exception of the WS topology, these results are common to all the networks, including the empirical ones. 

Moreover, Fig.~\ref{picture 1} reveals that the \textit{transitions from a polarized state to consensus are always sharper for the probabilistic model} than for the corresponding deterministic one, regardless of the topology. 

Interestingly, these two modifications of the original model compete, but their effects are not modified by their combination: one observes that a fuzzy neighbor selection \textit{always favors consensus, regardless of the type of influence function} (compare, for a fixed topology, model $HK_2$ with $HK_0$ which do not reinforce confirmation bias, and models $HK_3$ with $HK_1$, which do). Conversely, enhancing confirmation bias in the influence function \textit{always hinders consensus, regardless of the type of neighbor selection} (compare, for a fixed topology, models $HK_0$ with $HK_1$, which have a deterministic selection of neighbors, and models $HK_3$ with $HK_2$, which use a probabilistic one). 

While the size of the largest opinion cluster provides information on the degree of consensus achieved, its final opinion value provides insights into the extremism of the opinion adopted by the majority. As expected in the case of a homogeneous system with uniformly distributed initial opinions, consensus is reached at the middle of the opinion space (around $0.5$), for all the models and topologies (see details in Section S3 of the Supplementary Material).
  
Although these qualitative results are generally independent of the network considered, the detailed behavior of the order parameter may depend on the network structure. First, Fig.~\ref{picture 1} illustrates a phenomenon already observed in previous works~\cite{schawe_when_2021,perrier_phase_2024}: there is a clear difference between the behavior of the order parameter on purely random networks (ER, BA, and the empirical ones) on the one hand, and on the WS network on the other. For the latter, instead of a sharp transition, one observes a smooth crossover (the peaks of the variance vanish) for the models $HK_0$ and $HK_1$, as confirmed by a finite-size study. This result is related to the well-known fact of the absence of phase transition in one-dimensional lattices with short-range interactions (cf. standard results of the theory of phase transitions)~\cite{goldenfeld_lectures_2018}. The WS network consists of a slight randomization of a one-dimensional lattice (with a small number of shortcuts that might introduce some longer-range interactions), from which it retains the main characteristics. Although in Fig.~\ref{picture 1} the sharp transition seems to be re-established when the neighbors' selection is fuzzy (models $HK_2$ and $HK_3$), a finite size study shows that the variance diminishes with size (see Section S4 of the Supplementary Material).

Second, if we focus only on networks with purely random topology, either stylized network models (BA and ER) or empirical networks, Fig.~\ref{picture 1} shows that the order parameter, particularly for models with strict neighbor selection criterion ($HK_0$ and $HK_1$), behaves similarly for the two stylized networks, despite their different degree distributions. However, this behavior is different for empirical networks, which display strong community structure. In this case, for values of $\varepsilon$ in the consensus region, the value of $<S>$ is systematically smaller than for stylized  networks. Moreover, in this case, the growth of $<S>$ with $\varepsilon$ after the appearance of a giant cluster follows a linear trend. We have established that this is caused by the large number of agents with degree $k \approx 1$, present in the empirical networks (see Supplementary Materials, section S5).

\begin{figure*}[htbp!]
    \centering
    \includegraphics[width=\textwidth]{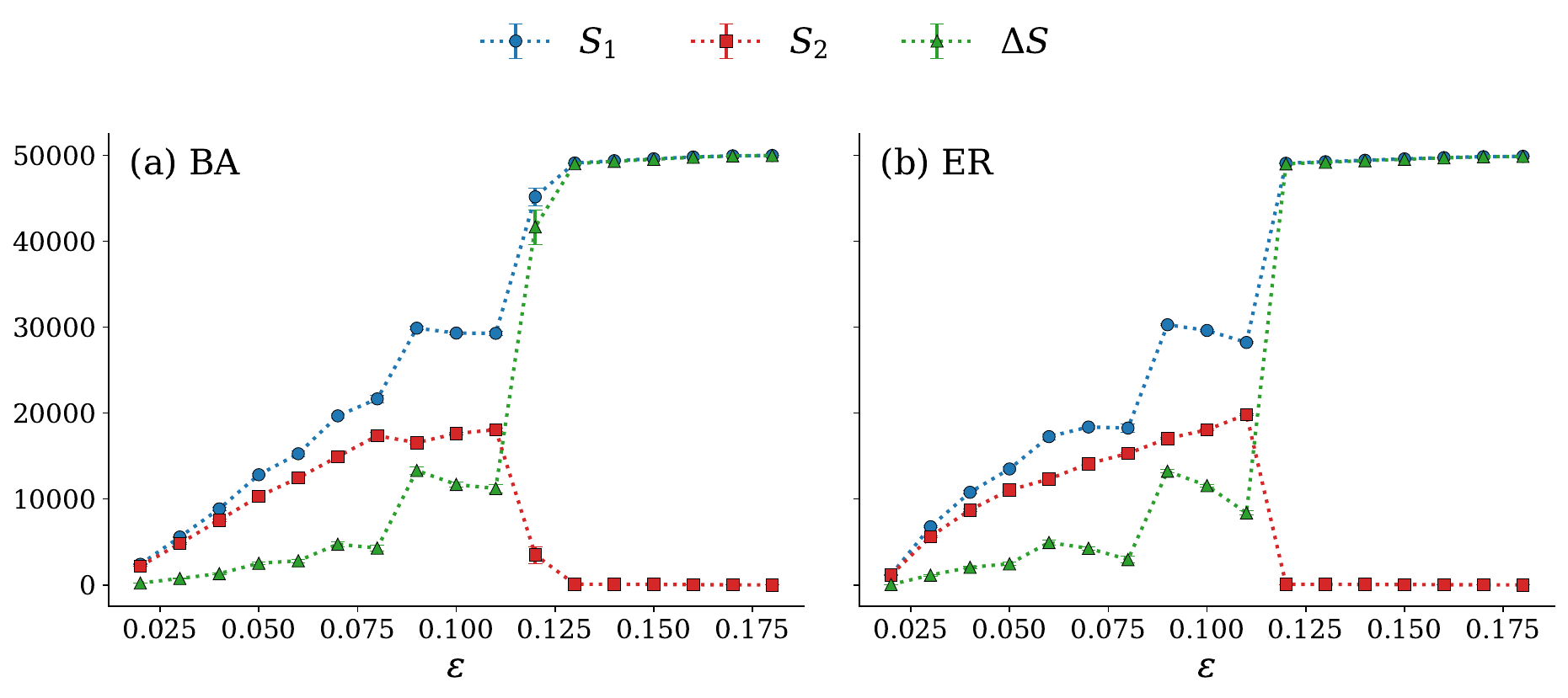}
    \caption{Sizes of the two largest opinion clusters, $S_1$ and $S_2$, and their difference $\Delta S = S_1 - S_2$ as a function of the confidence bound $\varepsilon$ for model $HK_3$ ($\beta=3$, uniform initial opinions), averaged over $100$ realizations for each $\varepsilon$ value. Left: Barabási–Albert (BA) network. Right: Erdős–Rényi (ER) network. In the plateau region, $\Delta S$ decreases as $S_2$ grows and $S_1$ slightly decreases, indicating a regime of competing clusters with comparable sizes prior to consensus. Dotted lines are only a guideline for the eye.
}
    \label{fig:cluster_dynamics}
\end{figure*}

Another striking difference between the empirical networks and the stylized models is the behavior of probabilistic models ($HK_2$ and $HK_3$), which show a kink or a well-defined plateau in the polarization region ($<S> \approx 0.5$) for the stylized networks, which is absent in the empirical ones. As in the polarization region there are two opinion clusters of comparable size, $S_1$ and $S_2$, for each sample, we show in Fig.~\ref{fig:cluster_dynamics} their behavior as a function of $\varepsilon$, along with that of their difference, $\Delta S = S_1 - S_2$.
 
In the stylized networks, for $\varepsilon$ values in the plateau region, there is a decreasing trend in the size difference between the two largest clusters, $\Delta S$. This indicates a state of competitive polarization in which two major opinion clusters resist merging until they eventually collapse into a single cluster, marking the end of the plateau. This phenomenon does not occur in the empirical networks, which exhibit strong heterogeneity in the $S_1$ and $S_2$ values across different realizations for the same $\varepsilon$.

We can quantify this variability by computing the Shannon entropy of the distribution of size differences of the two largest clusters, for each value of $\varepsilon$:

\begin{equation}
\centering
    H = - \sum_{b} p_b \log(p_b) \label{eq:shannon function}
\end{equation}

where $p_b$ corresponds to the probability of observing a value of $\Delta S$ associated with bin $b$. We found that the entropy of the empirical networks in the corresponding transition region is always higher than that of the stylized networks, thus confirming the greater variability in the sizes of the two dominant clusters. More details on these results can be found in section S6 of the Supplementary Material.

Since the topology of the empirical networks is fixed across realizations, a parsimonious hypothesis is to relate the observed heterogeneity in $\Delta S$ to the random initialization of opinions. Therefore, the initial opinions assigned to highly connected nodes play a crucial role. The empirical networks contain hubs with extremely large degrees, to which many low-degree nodes are attached. If such hubs are initially assigned different extreme opinions, they can hinder global consensus, leading to final states where $S_1$ and $S_2$ are comparable. Conversely, when the initial opinions of these hubs are closer to the center of the opinion space, they tend to promote consensus, resulting in $S_1 \gg S_2$ and, in turn, leading to the observed broad distribution of $\Delta S$ (see details in Section S6 of the Supplementary Material).

%============================
% Gaussian initalization
%============================
When comparing the outcomes of the different models for empirical networks with the patterns observed in real systems, we must take into account an important aspect of online networks that has been neglected so far. On the one hand, the models studied here consider a fixed network. On the other hand, the community structure observed in empirical retweet networks is the result of connections driven by homophily, as people typically decide to retweet other agents with whom they agree. Therefore, if we want to run our models on the empirical networks "as they are" at the time of capture, we should revise the initialization of opinions to account for the effect of homophily, as a uniform distribution of initial opinions seems quite unrealistic.

To take this into account, we simulate our models by replacing the random uniform initialization of the agents' opinions with the truncated Gaussian distribution described in Section~\ref{sec:opinion initialization}. This distribution mimics the main characteristics of our dataset, where we observe that some of the largest communities correspond to well-identified political orientations (see Section S7 of the Supplementary Material for some examples).

Fig.~\ref{fig:GI_comparison} shows the comparison of the order parameter $<S>$ as a function of $\varepsilon$ for the two different initializations and all the studied models on the GI network (results for the other empirical network are shown in Section S7 of the Supplementary Material). While the main properties of the models discussed above are independent of the initialization---fuzzy neighbor selection promotes consensus, while enhanced confirmation bias hinders it---, Gaussian initialization increases consensus for all models with respect to uniform initialization, especially for models with the strict selection criterion ($HK_0$ and $HK_1$). This is not surprising given that the pre-clustering of initial opinions of the largest communities allows for many more interactions from the beginning of the dynamics, promoting consensus. 
 
\begin{figure*}
    \centering
        \includegraphics[width=\textwidth]{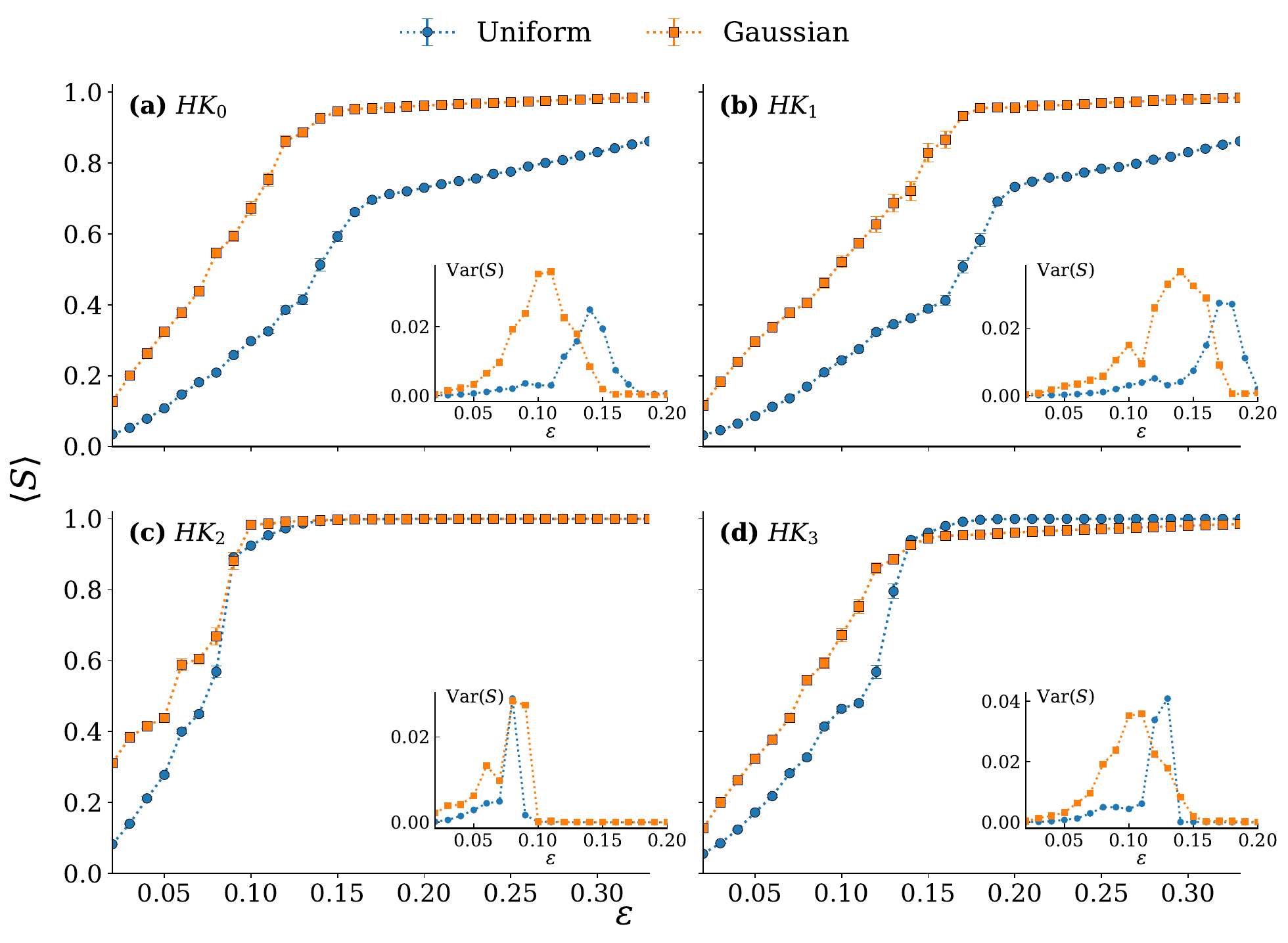}
    \caption{Order parameter $<S>$ as a function of $\varepsilon$ on the empirical GI network for the four models ($HK_0$–$HK_3$), comparing uniform and Gaussian initial conditions. Insets show the variance of $S$, used to estimate the critical threshold $\varepsilon_c$. Gaussian initialization systematically promotes consensus, shifting the transition to lower $\varepsilon$ values for models with strict neighbor selection ($HK_0$, $HK_1$), while having a weaker effect on models with fuzzy selection ($HK_2$, $HK_3$). Dotted lines are only a guideline for the eye.
}
    \label{fig:GI_comparison}
\end{figure*}

\begin{figure*}[htbp!]
    \centering
    \begin{subfigure}[b]{0.49\textwidth}
        \centering
        \includegraphics[width=\textwidth]{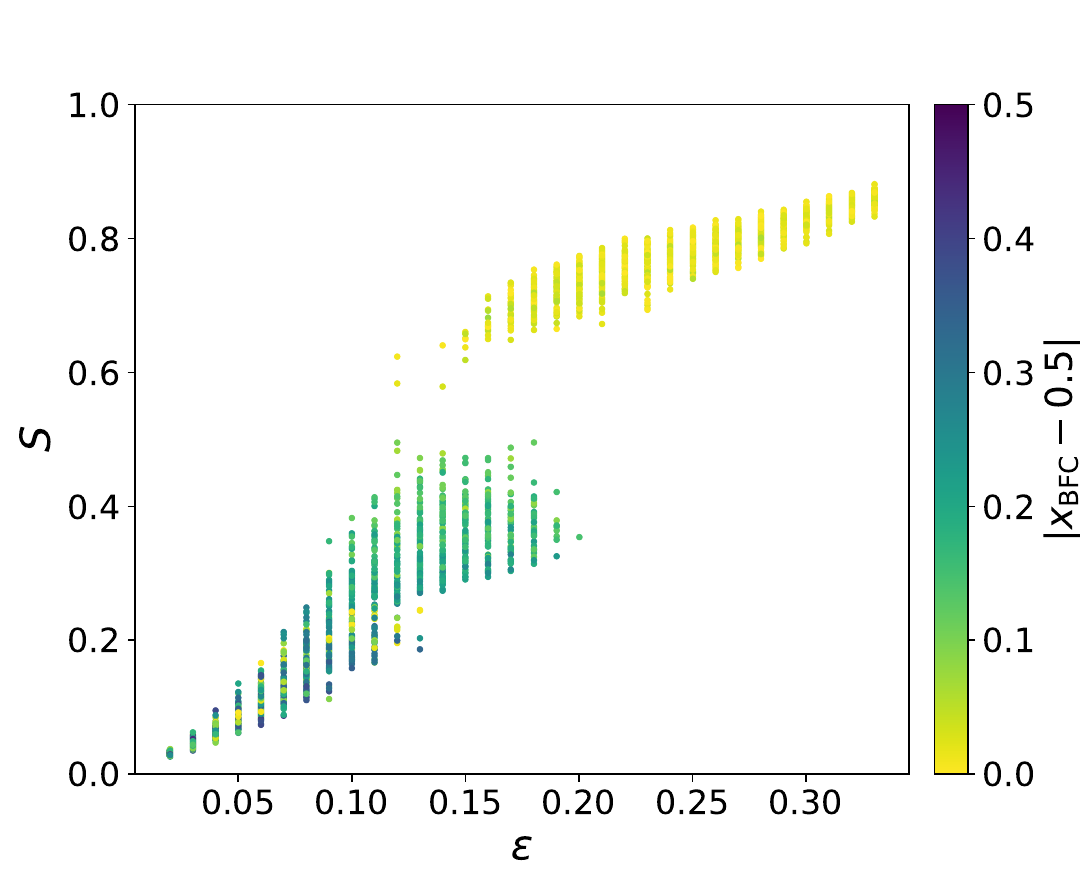} 
        \caption{Uniform Initialization}
        \label{fig:bfc_shift_uniform}
    \end{subfigure}
    \hfill 
    \begin{subfigure}[b]{0.49\textwidth}
        \centering
        \includegraphics[width=\textwidth]{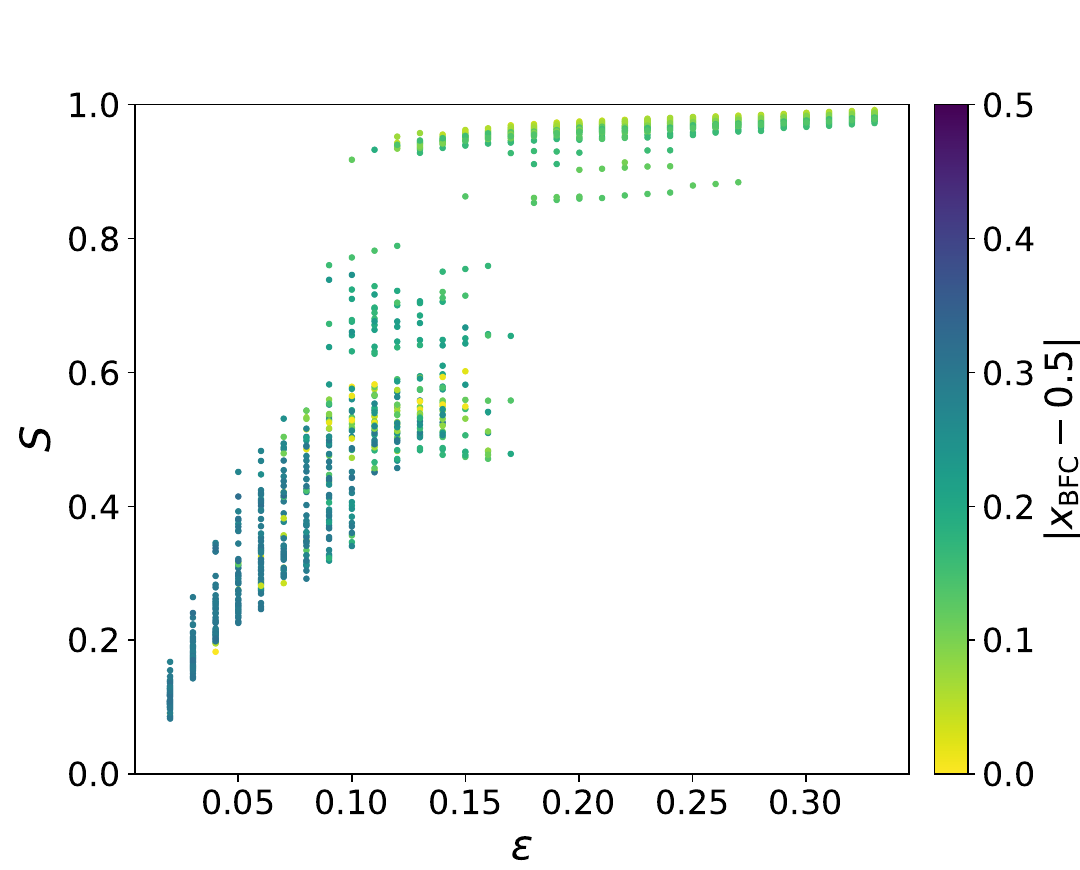} 
        \caption{Gaussian Initialization}
        \label{fig:bfc_shift_gaussian}
    \end{subfigure}
    \caption{Comparison of the largest final cluster (BFC) for model $HK_1$ on the GI network under uniform (a) and Gaussian (b) initial conditions. Each point represents one realization, showing the BFC size as a function of $\varepsilon$. Colors indicate the deviation of the final consensus opinion from the center, $|x_{\mathrm{BFC}} - 0.5|$. While uniform initialization leads to consensus around the central opinion, Gaussian initialization produces a systematic shift toward more extreme values, reflecting the influence of the initial community structure.}
    \label{fig:bcf_opinion_shift}
\end{figure*}

Another consequence of Gaussian initial conditions is that the final opinion of the dominant cluster in the consensus region can be shifted toward the extremes, even in a homogeneous system (there is a single $\varepsilon$ for all agents), given the interplay between initial conditions and community structure, as shown in Fig. \ref{fig:bcf_opinion_shift} for the $HK_1$ model on the GI network. This shift is a direct consequence of the fact that the largest initial topological community, which corresponds to the extreme-right leader \textit{Le Pen}, was initialized with a mean opinion around $0.8$. This illustrates how a realistic, community-based initial opinion structure can determine not only the size of the final consensus but, crucially, its opinion. Examples of trajectories for the two initializations in the consensus region can be found in Section S7 of the Supplementary Material.

\begin{figure*}[t]
    \centering
    \begin{subfigure}[b]{\textwidth}
        \centering
        \includegraphics[width=\textwidth]{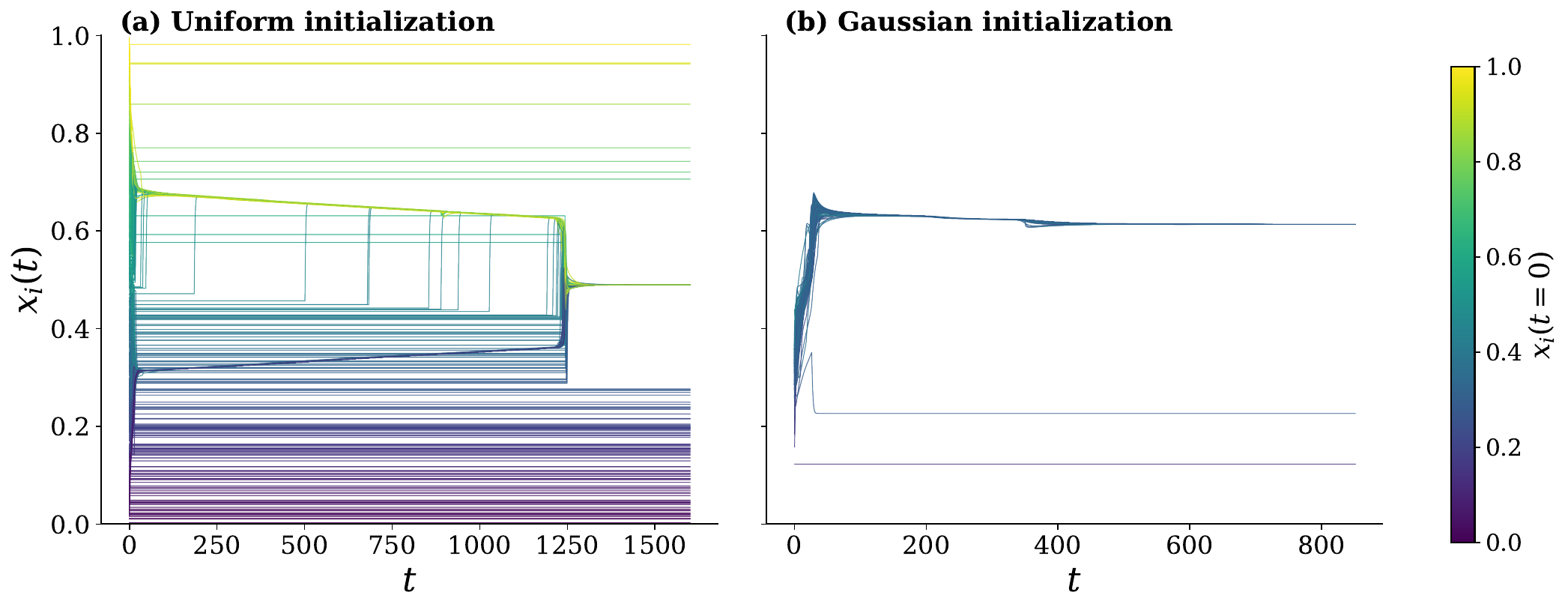}
    \end{subfigure}
    \vfill
    \begin{subfigure}[b]{\textwidth}
        \centering
        \includegraphics[width=\textwidth]{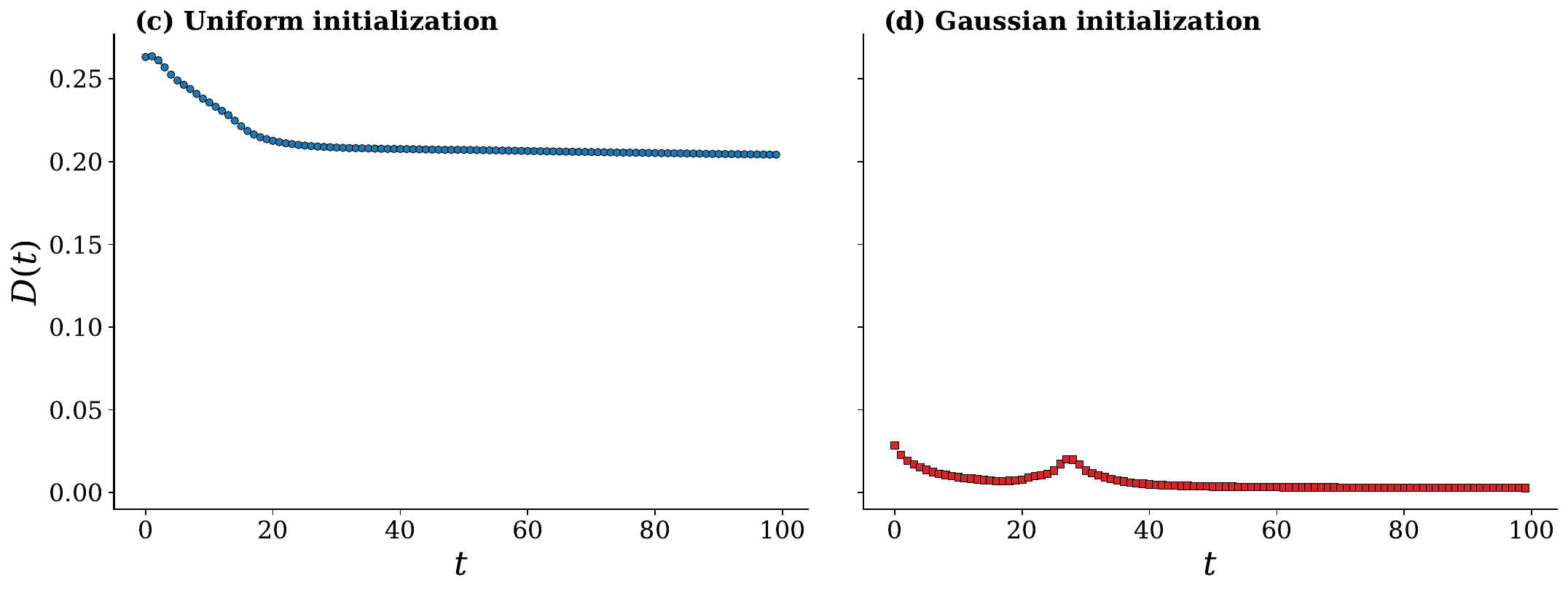}
    \end{subfigure}
    \caption{Time evolution of opinions within a representative community $C$ ($N_C=664$) on the GI network for model $HK_1$ at $\varepsilon=0.2$. Top row: Gaussian initialization. Bottom row: uniform initialization. Left panels show the opinion trajectories $x_i(t)$ of agents in $C$. Right panels show the evolution of the intra-community dispersion $D(t)$ (Eq.~\ref{eq:D_function}), in the first $100$ time steps of each realization. Under Gaussian initialization, a two-step dynamics is observed: a rapid decrease of $D(t)$ indicates fast local convergence within the community, followed by a global shift toward the consensus cluster (corresponding to the small peak). In contrast, uniform initialization shows no clear local consolidation, with a slower and more gradual decrease of $D(t)$, consistent with agents joining the global consensus individually rather than as a cohesive group.}
    \label{fig:community_dynamics}
\end{figure*}

Finally, we want to understand whether a community structure based on homophily influences not only the steady state but also the dynamics. Specifically, we want to understand \textit{the road to consensus}: whether it is achieved via a local convergence (communities consolidate internally first and merge afterwards) or by a direct global convergence (individual agents join the main opinion cluster independently, regardless of the community they belong to). To study this, we consider model $HK_1$, at $\varepsilon=0.2$, where consensus is achieved with both uniform and Gaussian initializations. In both cases, we focus on the dynamics of a specific community $C$, of size $N_C=664$, and we study the opinion evolution of its agents. In order to quantify the level of internal opinion convergence, we define the quantity $D(t)$:

\begin{equation}
\centering
    D(t) = \frac{\sum_{i \in C} |x_i(t) - \mu_C(t)|}{|C|} \label{eq:D_function}
\end{equation}

where $\mu_C(t)$ is the average opinion inside community $C$ at time $t$. $D(t)$ measures the evolution of the average opinion distance of agents within the community from their community mean. Figure~\ref{fig:community_dynamics} presents the opinion trajectories and the evolution of $D(t)$ for the community $C$ under both uniform and Gaussian initialization.

\FloatBarrier

\FloatBarrier

Under Gaussian initialization, we observe a clear two-step process: agents first achieve local convergence, concentrating rapidly around their community's average opinion. This is evidenced by the initial drop in $D(t)$. A subsequent increase in $D(t)$ marks the moment when the entire community shifts globally to join the main opinion cluster, driven by agents sharing connections with other communities. In sharp contrast, the uniform initialization shows no preliminary local convergence. Due to the high initial opinion distance, reflected in higher initial $D(t)$ values, interactions are limited to agents whose opinion difference is within the bounded confidence $\varepsilon$. Consequently, the $D(t)$ curve undergoes only a slow, gradual decrease. This behavior indicates that agents join the global cluster individually rather than as a cohesive community unit. These observations confirm that the initial opinion distribution dictates whether consensus is reached through community-level consolidation or through a process resembling the network's overall global dynamics.

\section{Discussion}\label{sec:discussions}

This work provides a systematic and comprehensive perspective on the effect of introducing some widespread characteristics of real systems into the classic HK model in networked societies. Specifically, we investigate the role of a probabilistic selection of the interacting neighbors, accounting for the fact that agents do not have infinite precision in evaluating others' opinions, as well as the role of introducing a heterogeneous confirmation bias in the influence function. 

The most general conclusion derived from the simulations of the modified models ($HK_1$, $HK_2$, and $HK_3$) is the qualitative robustness of their effects on consensus. Irrespective of the network topology (stylized or realistic) and regardless of the initial opinion distribution for the empirical networks (uniform or Gaussian), the introduction of fuzzy selection of interacting neighbors \textit{facilitates consensus} by reducing the $\varepsilon$ value that leads to it. This effect of the "imperfect knowledge" about the actual value of others' opinions, which is required to determine the set of neighbors that intervene in each local interaction, has  also been observed for a Deffuant-Weisbuch model where fuzziness in the selection of the interacting pairs was introduced in a very different form~\cite{steiglechner_noise_2024}. 

Moreover, for empirical networks the probabilistic selection of neighbors makes the transition between consensus and polarization much sharper, leading to a narrower critical region in $\varepsilon$, as shown by the $HK_0$ and $HK_2$ curves in Fig.~\ref{picture 1}(d). Although we can't argue for the discontinuous nature of this transition without a finite-size analysis, which is excluded in the case of a specific empirical network, the observed decrease of the transition region for fuzzy neighbor selection is important for understanding the evolution of opinions in real settings, such as online social networks. A system featuring sharp transitions from high values of $<S>$ implies that a slight reduction in people's open-mindedness could result in a sudden and drastic fragmentation of the opinion space if the system is initially near $\varepsilon_c$.  
It is for example known that confirmation bias is enhanced by exposure to toxic content \cite{knobloch-westerwick_confirmation_2020} and that some parametrizations of social network recommender systems (like engagement maximization) can increase significantly an exposure of their users to toxic content \cite{chavalarias_can_2024, bouchaud_crowdsourced_2023}, thus decreasing the $\varepsilon$ of the society. Consequently, as the deterministic selection of neighbors by a sharp threshold is unrealistic, $HK_2$ shows that a sharp transition from consensus to polarization can be expected following a small decrease of the confidence bound that parametrizes the neighbors' selection probability.

On the other hand, the influence function with weighted updates systematically \textit{hinders consensus}, because it increases the minimum value of $\varepsilon$ that leads to consensus. These results show how the effect of an enhanced confirmation bias in the influence function is partially counterbalanced by the unavoidable uncertainty affecting the choice of the interacting neighbors.

A previous work~\cite{grauwin_opinion_2012} studied a version of the DW model with a fuzzy  neighbor selection, in a manner similar to that of $HK_2$ described here. However, that study considered a fully-mixed population and very small stylized networks. Our results for $HK_2$ extend their findings to the HK model and to much larger networks (over two orders of magnitude larger), including empirical ones, thus showing the key role that the uncertainty in the evaluation of others' opinions plays in bounded confidence models.  

Our analysis further clarifies the role played by some properties of real network topologies, such as strong heterogeneity in the degree distribution and community structure. For example, we observed that for models with strict neighbor selection ($HK_0$ and $HK_1$), the final largest opinion cluster, in the consensus regime, is significantly smaller in the GI empirical network compared to synthetic networks of comparable size, connectivity and degree distribution, such as the BA network. Our results reveal that this difference is driven by details of the network’s degree distribution. In fact, although both networks have strongly heterogeneous distributions, the results can be understood by noting the higher prevalence of low-degree nodes in the empirical network. The very few connections of these low-degree agents prevent them from shifting their opinions toward that of the dominant opinion cluster. This result highlights a crucial point concerning the models of empirical networks: the specific architecture of the degree distribution may be determinant. In this case, the large amount of low-degree nodes creates pockets of resistance to convergence.

Furthermore, a strong difference is observed between the results obtained for the fully random stylized topologies (ER and BA) and those observed in empirical networks. The former exhibit a region of $\varepsilon$ characterized by a \textit{polarization plateau} that is absent in the latter. The mechanism leading to such a plateau is the existence of two clearly identified opinion clusters of similar  size that emerge in the steady state. For the empirical networks with uniform initial conditions we observe that the distribution of the differences in size between the largest and second-largest opinion clusters is broader than for the stylized networks and clearly depends on the specific initial opinions of the hubs belonging to different communities.

Another contribution of this study lies in exploring community-level consensus dynamics in empirical networks by adapting the opinion initialization to a more realistic setting that takes into account the role of homophily in the very formation of communities in social networks. Unlike the standard uniform initialization of the model, which leads to a global convergence mechanism, the introduction of Gaussian initialization leads to a robust two-phase process: convergence takes place first locally and eventually becomes global. This suggests that in realistic social settings where opinions are pre-clustered within communities, although with a finite dispersion, global consensus is not merely an aggregation of individual actions but rather the merging of cohesive community entities. The rapid internalization of the average opinion within a community, before the entire community shifts towards the main cluster, driven by boundary agents, illustrates how social cohesion mediates the dynamics of mass opinion adoption.

\section{Conclusions}\label{sec:conclusions}

This work shows the dominant role that realistic properties of social actors, such as the uncertainty in the evaluation of others' opinions or the enhancement of confirmation bias in the influence function, have in the outcomes of the HK model. The study of the influence of these ubiquitous properties in well-known opinion dynamics models is fundamental to developing general models that could be compared to empirical systems. We established that the qualitative effects of the fuzzy selection mechanism (which promotes consensus) and of enhanced confirmation bias in the influence function (which inhibits it) are robust and invariant with respect to topology and initialization, with the former having a dominant effect. Another insight is the discovery of a two-stage convergence mechanism when considering a real-world network: the more realistic Gaussian initialization triggers a convergence that is first local within communities and then global, suggesting that community cohesion is an essential factor in mass consensus in pre-clustered social contexts. 

Because of its interesting results, this work constitutes a further step in the ongoing endeavor to bridge the gap between theoretical models and empirical studies, which must be pursued further. For example, the current model assumes static and undirected networks. Future research should address these limitations by considering directed networks, in order to study non-reciprocal influence, and by exploring the mechanisms of network formation, instead of starting from already existing empirical networks, thus offering a more complete view of the co-evolution of opinions and social structure.

Furthermore, the  extraction of the probability of interaction from empirical networks~\cite{chavalarias_can_2024}  allows us to simulate a probabilistic opinion model without a confidence bound, adapted to networks with strong community structure when opinions are pre-clustered according to some political affiliation, as observed in political campaigns. This work is currently in progress.

\bibliography{sample}

\section*{Author contributions statement}
CG developed and ran the codes, analyzed the results, and prepared the manuscript. LH designed the research, discussed the results, and corrected the manuscript. DC discussed research lines, provided the empirical data, and participated in the discussion of results. All authors contributed to the final version of the manuscript.

%\section*{Additional information}

% per Arxiv
\clearpage
\includepdf[pages=-]{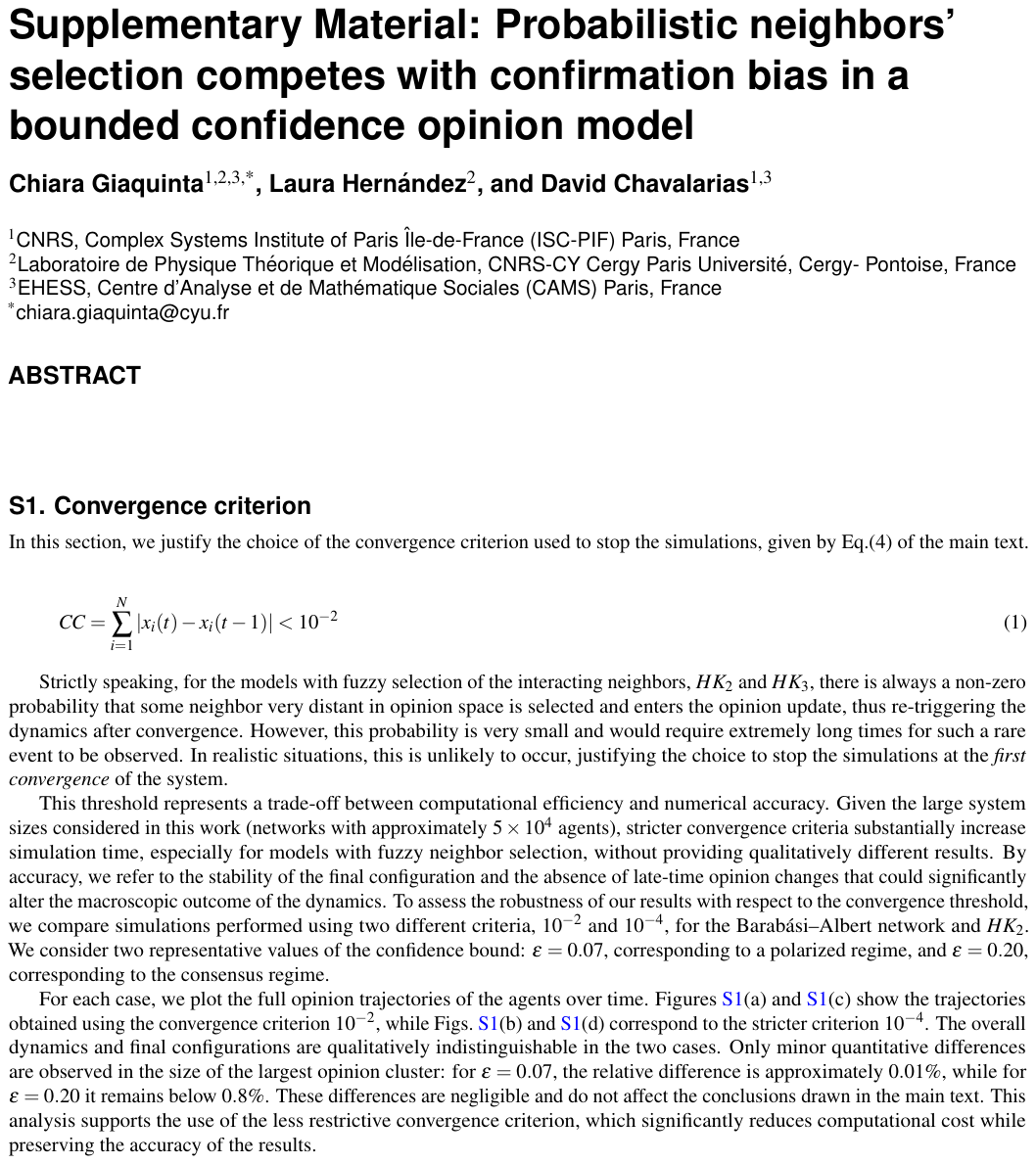}

\end{document}

\flushbottom
\maketitle
\thispagestyle{empty}

%\noindent Please note: Abbreviations should be introduced at the first mention in the main text – no abbreviations lists. Suggested structure of main text (not enforced) is provided below.

\renewcommand{\thefigure}{S\arabic{figure}}
\renewcommand{\thetable}{S\arabic{table}}

\section*{S1. Convergence criterion}

In this section, we justify the choice of the convergence criterion used to stop the simulations, given by Eq.(4) of the main text.

\begin{equation}
    CC = \sum_{i=1}^N |x_i(t) - x_i(t-1)| < 10^{-2}
\end{equation}

Strictly speaking, for the models with fuzzy selection of the interacting neighbors, $HK_2$ and $HK_3$, there is always a non-zero probability that some neighbor very distant in opinion space is selected and enters the opinion update, thus re-triggering the dynamics after convergence. However, this probability is very small and would require extremely long times for such a rare event to be observed. In realistic situations, this is unlikely to occur, justifying the choice to stop the simulations at the \textit{first convergence} of the system.

This threshold represents a trade-off between computational efficiency and numerical accuracy. Given the large system sizes considered in this work (networks with approximately $5 \times 10^4$ agents), stricter convergence criteria substantially increase simulation time, especially for models with fuzzy neighbor selection, without providing qualitatively different results. By accuracy, we refer to the stability of the final configuration and the absence of late-time opinion changes that could significantly alter the macroscopic outcome of the dynamics. To assess the robustness of our results with respect to the convergence threshold, we compare simulations performed using two different criteria, $10^{-2}$ and $10^{-4}$, for the Barabási–Albert network and $HK_2$. We consider two representative values of the confidence bound: $\varepsilon = 0.07$, corresponding to a polarized regime, and $\varepsilon = 0.20$, corresponding to the consensus regime.

\begin{figure*}[htbp]
    \centering
    \includegraphics[width=\textwidth]{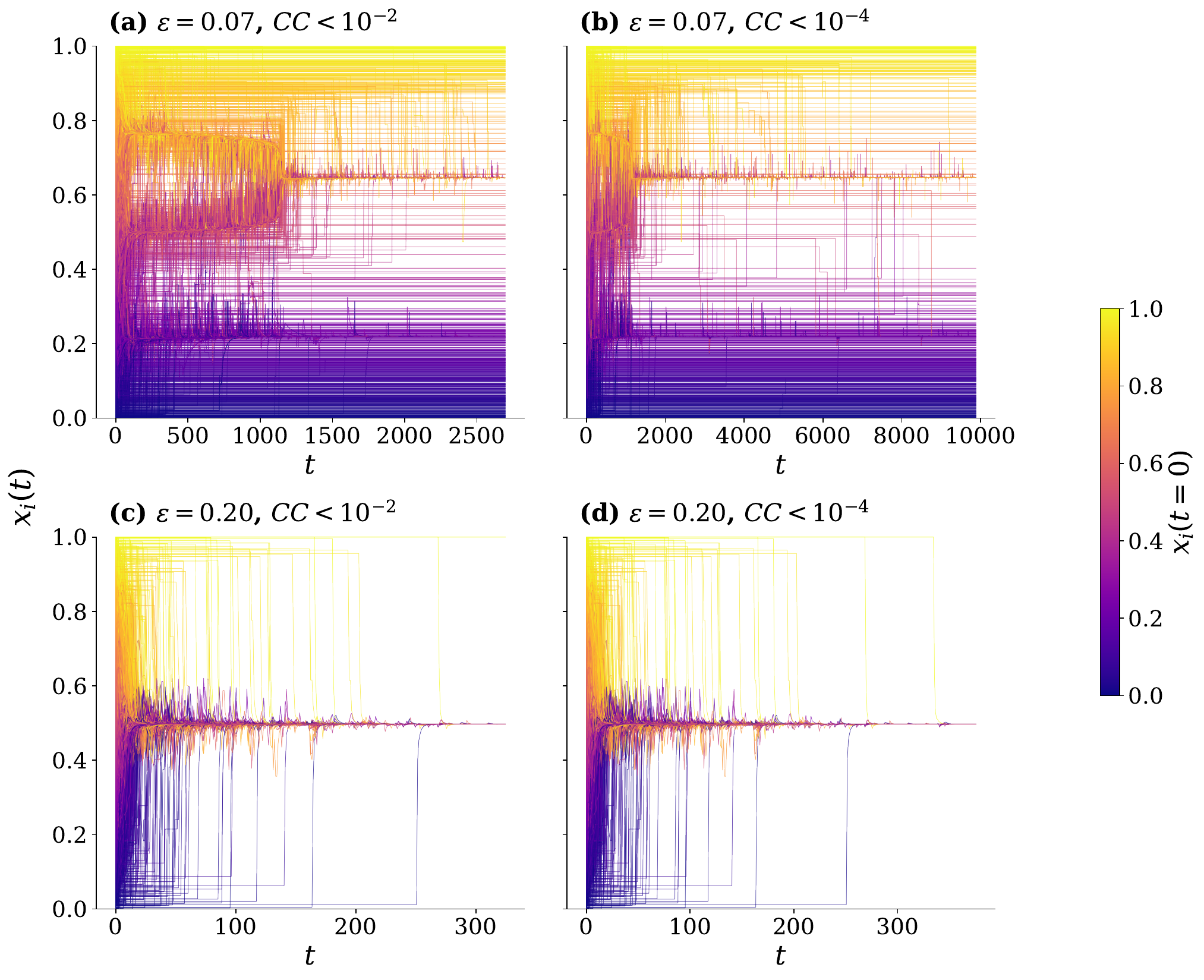} 
    \caption{Opinion trajectories for model $HK_2$ on a Barabási–Albert network ($N = 50000$), comparing two convergence criteria and two values of the confidence bound $\varepsilon$. Only a random sample of $15\%$ of agents is represented, for visualization purposes. Top row: $\varepsilon = 0.07$ (polarization regime). Bottom row: $\varepsilon = 0.20$ (consensus regime). Left column: convergence criterion $CC < 10^{-2}$. Right column: stricter convergence criterion $CC < 10^{-4}$. The trajectories and final configurations are qualitatively similar in all cases, indicating that the looser convergence criterion does not affect the macroscopic outcome of the dynamics, while strongly reducing the computational cost.}

    \label{fig:convergence_criterion}
\end{figure*}

For each case, we plot the full opinion trajectories of the agents over time. Figures~\ref{fig:convergence_criterion}(a) and \ref{fig:convergence_criterion}(c) show the trajectories obtained using the convergence criterion $10^{-2}$, while Figs.~\ref{fig:convergence_criterion}(b) and \ref{fig:convergence_criterion}(d) correspond to the stricter criterion $10^{-4}$. The overall dynamics and final configurations are qualitatively indistinguishable in the two cases. Only minor quantitative differences are observed in the size of the largest opinion cluster: for $\varepsilon = 0.07$, the relative difference is approximately $0.01\%$, while for $\varepsilon = 0.20$ it remains below $0.8\%$. These differences are negligible and do not affect the conclusions drawn in the main text. This analysis supports the use of the less restrictive convergence criterion, which significantly reduces computational cost while preserving the accuracy of the results.

\FloatBarrier

\FloatBarrier

\section*{S2. Description of empirical networks (GI and GP)}

\begin{figure*}[t]
    \centering
    \begin{subfigure}[b]{0.48\textwidth}
        \centering
        \includegraphics[width=\textwidth]{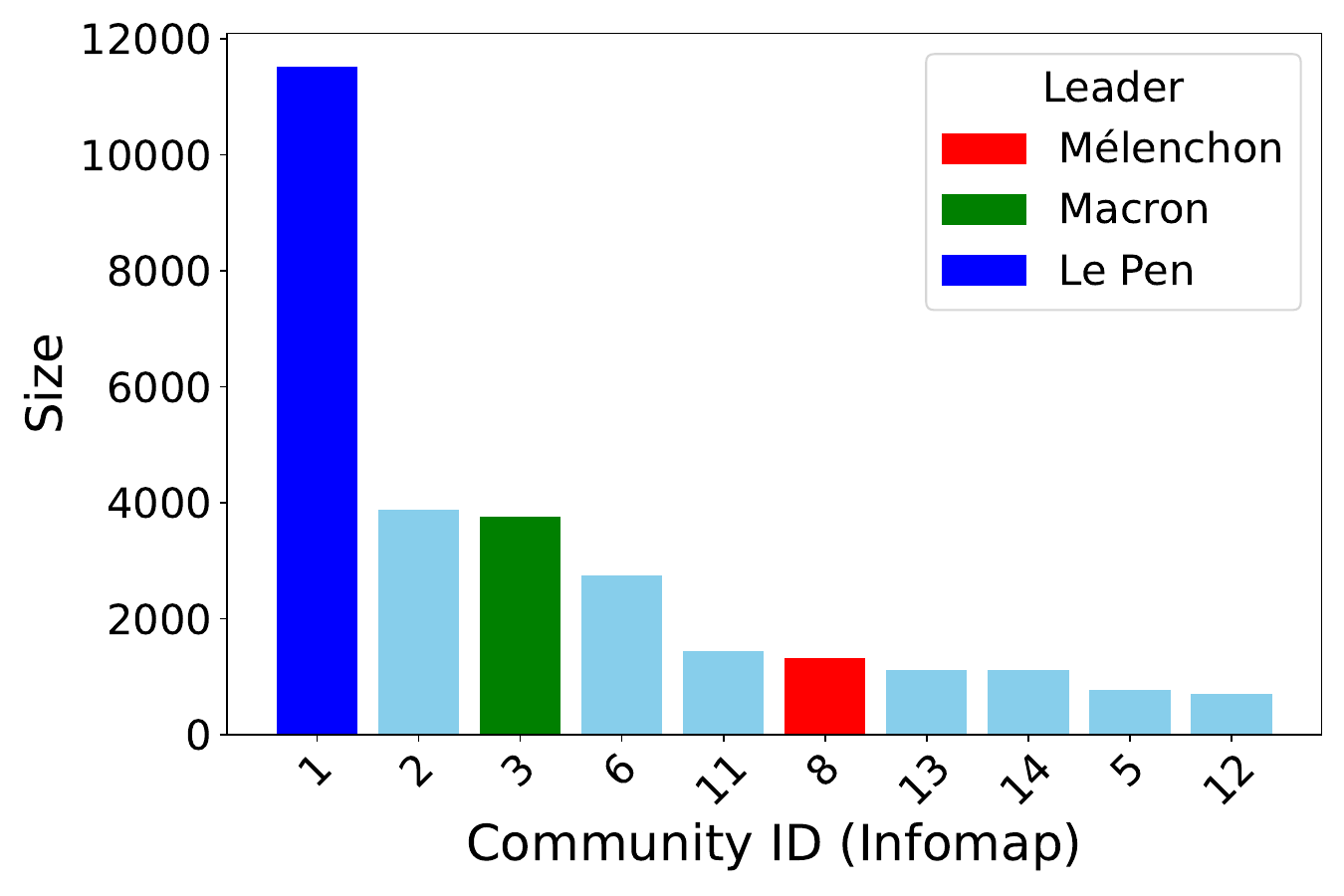} 
        \caption{GI}
        \label{fig:GI_comm}
    \end{subfigure}
    \hfill 
    \begin{subfigure}[b]{0.48\textwidth}
        \centering
        \includegraphics[width=\textwidth]{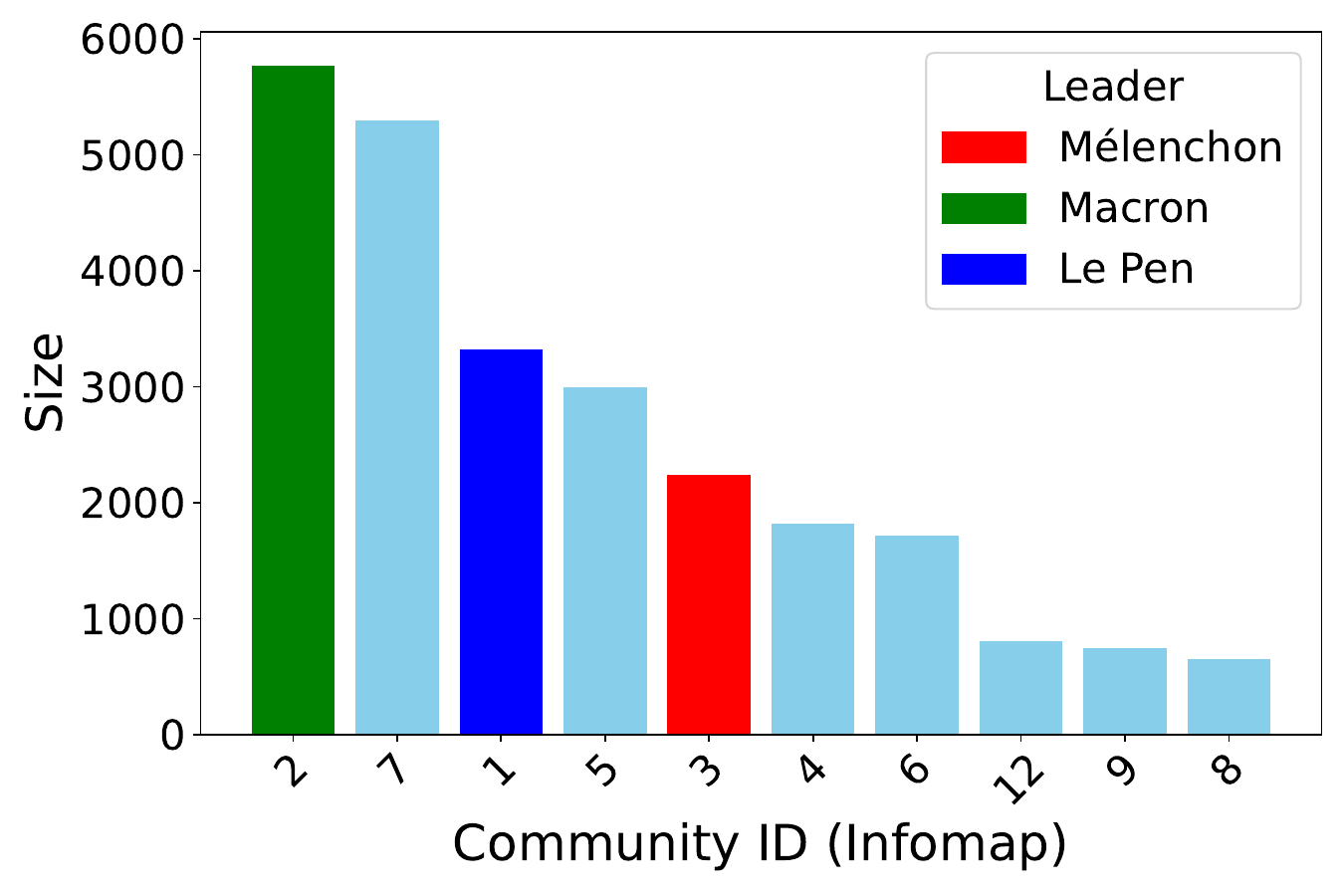} 
        \caption{GP}
        \label{fig:GP_comm}
    \end{subfigure}
    \caption{Size distribution of the largest communities detected with the Infomap algorithm in the GI (left) and GP (right) networks. The three dominant political communities in the 2017 French political landscape are highlighted with colors (red for Mélenchon, blue for Le Pen and green for Macron).}
    \label{fig:infomap_comm}
\end{figure*}

The real-world Network GI discussed in the main text was obtained from the Politoscope Project data~\cite{noauthor_projet_nodate}. Politoscope collected Twitter data (before the platform became X) focusing on French politics from 2016 to 2023. From these data, retweet networks can be built by distinguishing different discussion topics. The GI network discussed in the main text ($N=53002$, $<k> = 6.3$ after pre-processing), corresponds to the topic identified as \textit{"Immigration and Multiculturalism"} in France in 2017. For the sake of comparison, we studied a different empirical network (referred to as GP) that, after pre-processing, has comparable size ($N=42951$) and average degree $<k>=3.8$, and corresponds to the topic identified as \textit{"Foreign Politics"}. The obtained weighted and directed networks underwent several pre-processing steps: edges with weight one were excluded as noise, and the graphs were reduced to their giant connected components. They were then converted into unweighted and undirected networks. As discussed in the main text, we treat the resulting undirected networks as proxies for the underlying social interaction structure rather than for the directed diffusion of information on Twitter. Community detection was performed on these networks using the Infomap algorithm~\cite{edler_mapequation_2026}. Fig.~ \ref{fig:infomap_comm} shows the size distribution of their largest communities, highlighting the three communities corresponding to the main political leaders of that period according to their vote shares in the election (Macron, Le Pen, Mélenchon). 

\FloatBarrier

\FloatBarrier

\section*{S3. Beyond the order parameter: the scatterplots of independent realizations}

\begin{figure*}[htbp]
    \centering
    \includegraphics[width=\textwidth]{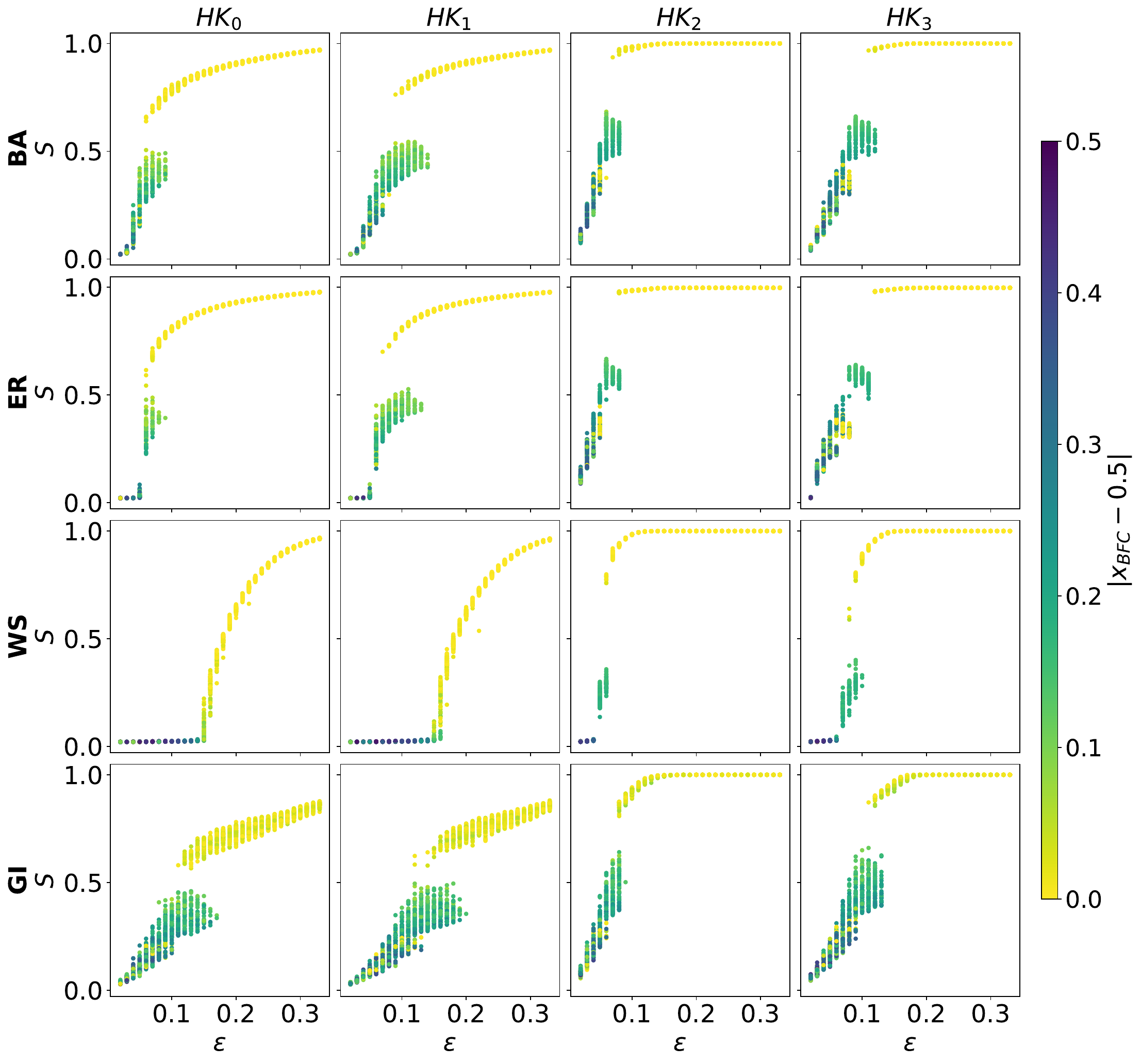} 
    \caption{Scatterplots of $S$ (largest opinion cluster size, normalized by network size) for uniform opinion initialization. Each row corresponds to a different network topology (Barabási--Albert, Erdős--Rényi, Watts--Strogatz, and the empirical GI network), and each column corresponds to a different model. Each point represents one independent realization of the initial conditions. Points are color-coded according to the final opinion value of the largest opinion cluster. Consensus is consistently reached near the central opinion value.}
    \label{fig:scatterplots_uniform}
\end{figure*}

As a complement to the plots of the order parameter $<S>$ as a function of $\varepsilon$ discussed in the main text, we present here the scatterplots of the sizes of the largest opinion clusters $S$ obtained in each individual realization, along with the opinion of that cluster in the steady state. This allows us to directly visualize the variability across independent runs. In Figure~\ref{fig:scatterplots_uniform}, we report these results for each model and network topology for uniform opinion initializations, when applicable.

We observe that consensus is consistently reached near the center of the opinion space (around $0.5$), as expected in the case of homogeneous systems with uniformly distributed initial opinions.

\section*{S4. Finite-size analysis of the transition in Watts–Strogatz networks}

In this section, we provide additional evidence supporting the absence of a true phase transition in Watts–Strogatz (WS) networks, as discussed in the main text. To this end, we analyze the behavior of the order parameter $<S>$ and its variance $Var(S)$ as a function of the confidence bound $\varepsilon$ for different system sizes ($N = 2,000$ to $50,000$), focusing on models $HK_0$, $HK_1$ and $HK_2$.

\begin{figure*}[htbp]
    \centering
    \includegraphics[width=\textwidth]{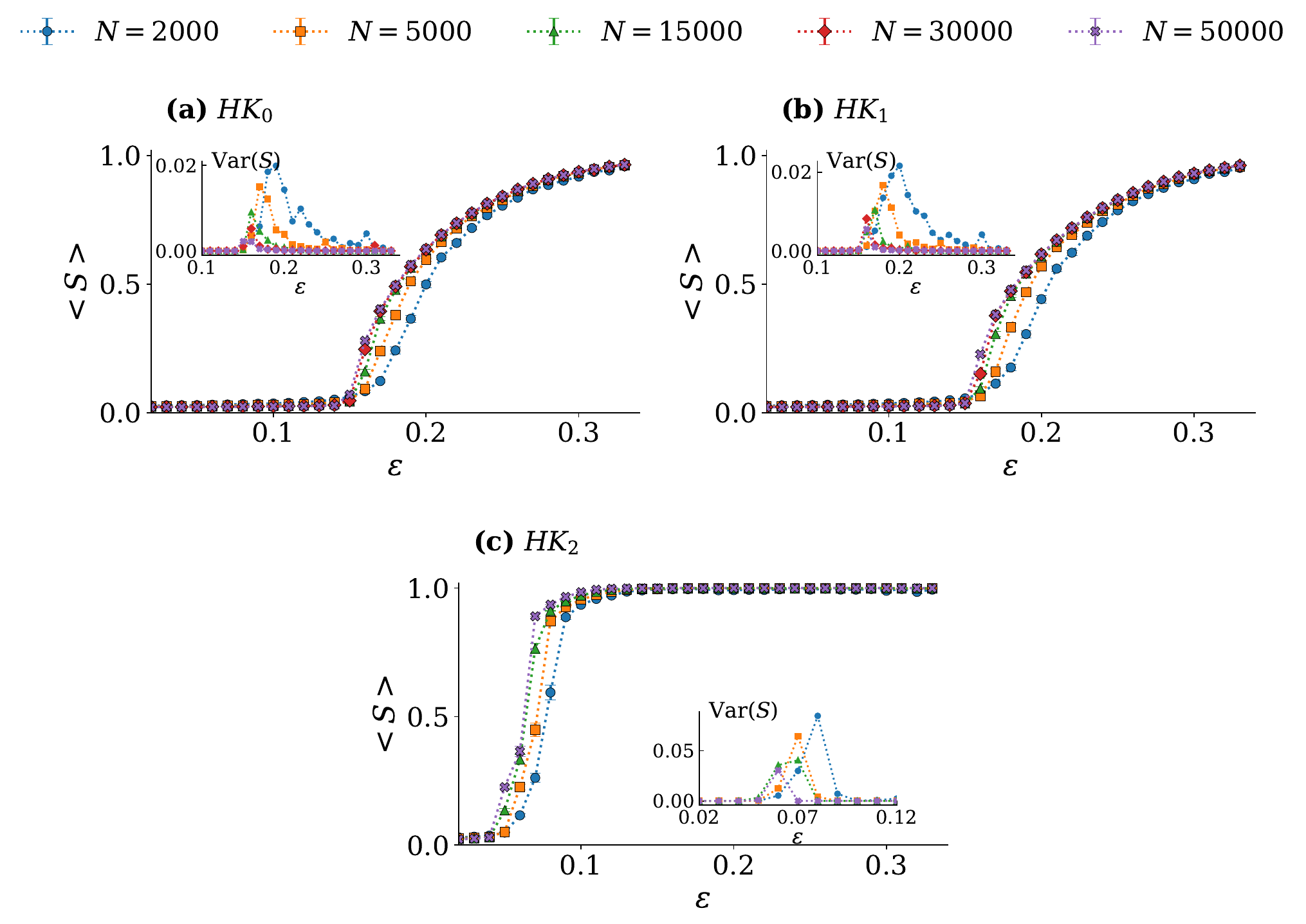} 
    \caption{Finite-size analysis of the order parameter $<S>$ as a function of $\varepsilon$ for Watts–Strogatz (WS) networks, for different system sizes $N$, in the $HK_0$, $HK_1$ and $HK_2$ ($\beta=3$) models. Insets show the corresponding variance $Var(S)$, used to identify the critical threshold $\varepsilon_c$. In all cases, the transition from fragmentation to consensus appears as a smooth crossover rather than a sharp phase transition. While $HK_2$ exhibits a steeper transition that may resemble a critical threshold for finite system sizes, no divergence of $Var(S)$ is observed. Instead, the peak seems to vanish as $N$ increases, indicating that the apparent critical behavior is a finite-size effect. These results support the absence of a true phase transition in WS networks, consistent with their quasi-one-dimensional structure and predominantly short-range interactions.}
    \label{fig:finite-size ws}
\end{figure*}

As shown in Figs.~\ref{fig:finite-size ws}(a) and \ref{fig:finite-size ws}(b), the transition from fragmented to consensus states appears as a smooth crossover rather than a sharp phase transition. This effect becomes more pronounced as the system size increases, with the $<S>$ curve progressively smoothing out with increasing $N$. A qualitatively similar behavior is also observed for $HK_2$ (in Figure \ref{fig:finite-size ws}(c)), where the transition appears significantly steeper and may resemble a critical threshold for finite system sizes. 

The insets of Fig.~\ref{fig:finite-size ws} show $Var(S)$, which is typically used to identify the critical point $\varepsilon_c$ in systems exhibiting a phase transition. For small system sizes, a peak in the variance is observed, suggesting critical behavior. However, as $N$ increases, the height of the peak decreases. This behavior indicates that the observed transition is a finite-size effect and that no true critical point exists in the thermodynamic limit. A similar trend is observed for $HK_2$, despite the sharper transition in $<S>$, confirming that the apparent criticality is a finite-size effect rather than a genuine phase transition.

This result is consistent with the known properties of one-dimensional systems with short-range interactions, which do not exhibit phase transitions. The WS network, being constructed from a regular lattice with a small fraction of random shortcuts, retains strong local structure and high clustering, effectively behaving as a quasi-one-dimensional system. Interestingly, this absence of a sharp transition contrasts with the behavior observed in Erdős–Rényi and Barabási–Albert networks, where a well-defined phase transition is recovered.

\section*{S5. Final opinions vs agents' degree}

Figures~2(a) and 2(d) of the main text showed that, considering values of the confidence bound $\varepsilon$ in the consensus regime (region of large values of $<S>$), empirical networks exhibit a smaller consensus cluster than stylized random networks. Moreover, for the deterministic models and unlike stylized networks, empirical networks show a region of a "quasi-linear" increase of $<S>$ with  $\varepsilon$.

To understand this fact, we compare the BA structure with the empirical networks because, unlike the ER network, they both have very heterogeneous degree distributions. However, the main striking difference between them is the community structure present in the empirical networks and absent from the BA one. Let us define as \textit{insider} an agent whose final opinion belongs to the largest opinion cluster, and as \textit{outsider} otherwise.  
By examining a specific realization in the consensus regime ($\varepsilon > 0.2$, $HK_1$), we observe that the largest cluster size is $S \approx 0.92$ for the BA network, but only $S \approx 0.74$ for the GI network. Our hypothesis is that, although both degree distributions are very heterogeneous, the observed difference is driven by a higher prevalence of low-degree nodes in the GI network than in the BA network. To test this hypothesis, we analyzed the final opinion of each agent as a function of its degree $k$. 

\begin{figure*}[htbp!]
    \centering
    \begin{subfigure}[b]{\textwidth}
        \centering
        \includegraphics[width=0.9\textwidth]{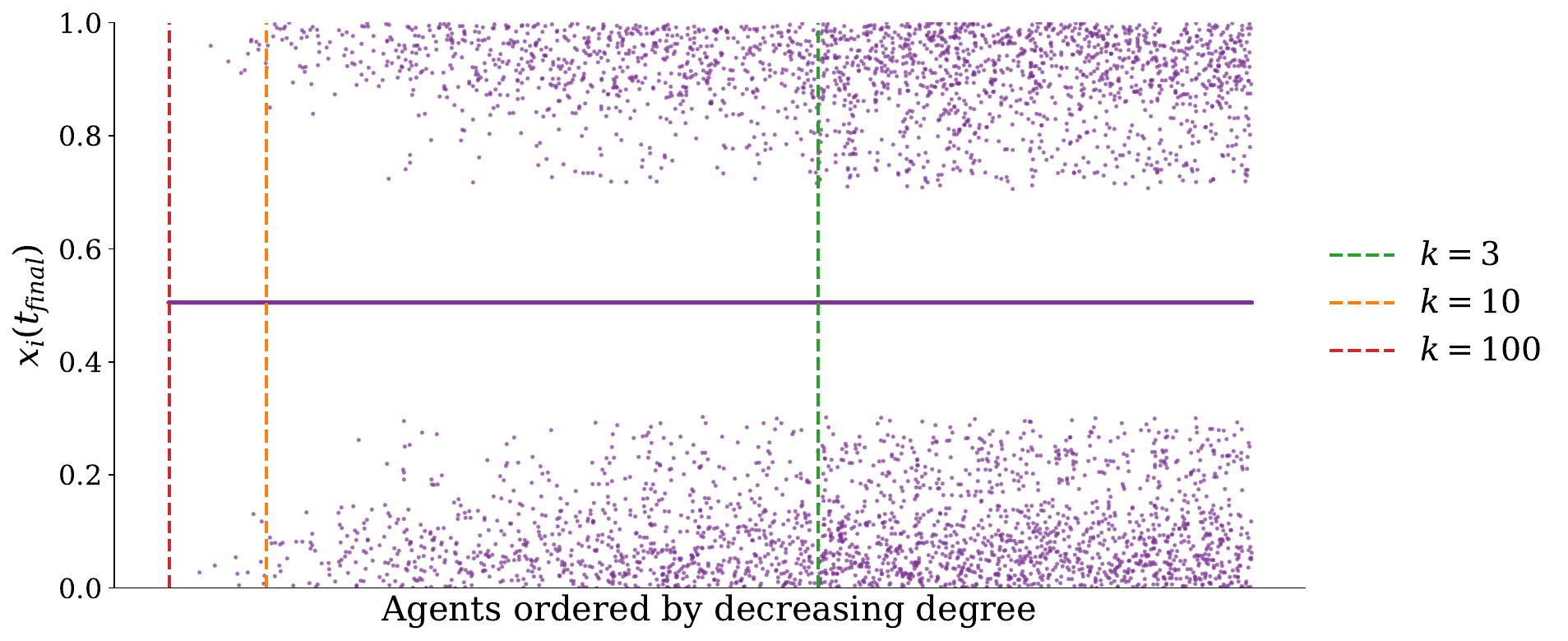}
        \caption{Barabasi-Albert (BA)}
        \label{fig:BA}
    \end{subfigure}
    \vfill
    \begin{subfigure}[b]{\textwidth}
        \centering
        \includegraphics[width=0.9\textwidth]{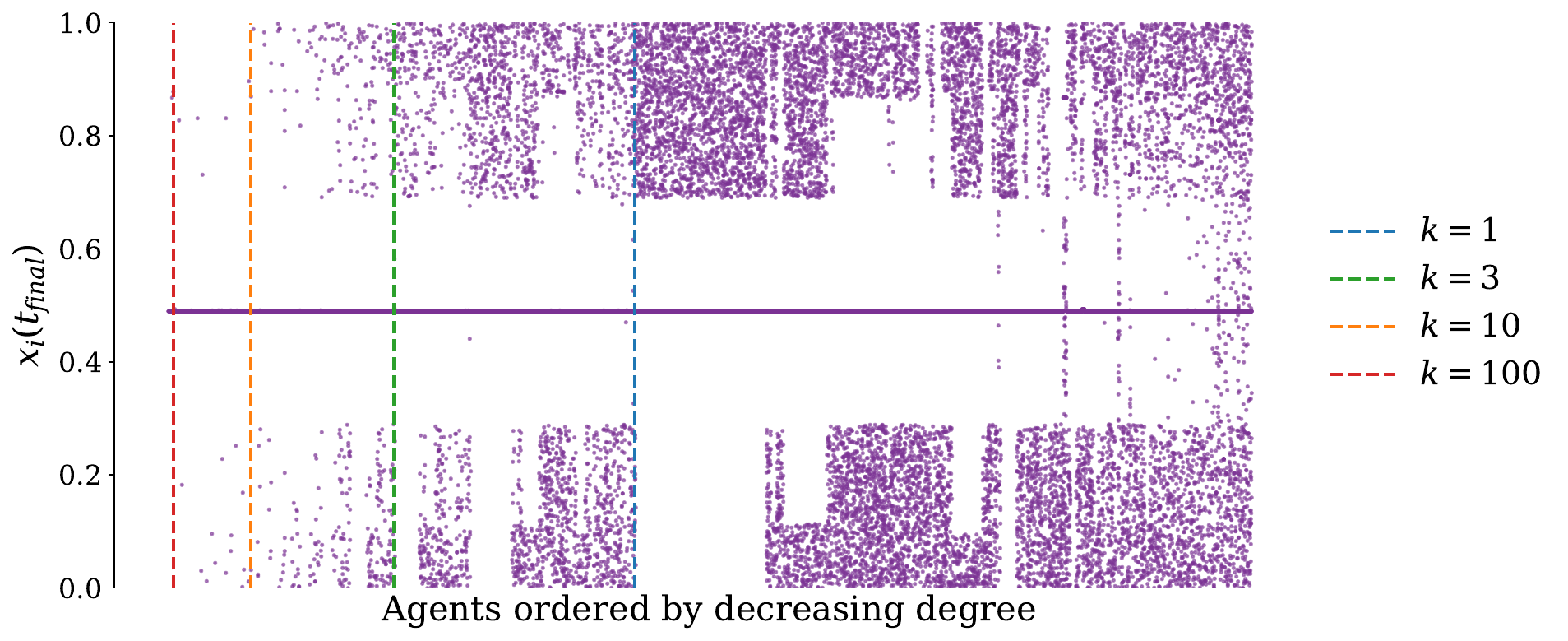}
        \caption{Real Network (GI)}
        \label{fig:GI}
    \end{subfigure}
    \caption{Final opinions $x_i(t_{final})$ as a function of node degree $k$ for a single realization of model $HK_1$ ($\varepsilon=0.2$, $\beta=3$, uniform initial opinions). Nodes are ordered by decreasing degree along the horizontal axis. Vertical dotted lines separate degree classes ($k \leq 1$, $1 < k \leq 3$, $3 < k \leq 10$, $10 < k \leq 100$, $k > 100$). Top: Barabási–Albert (BA) network. Bottom: empirical (GI) network. While both networks display heterogeneous degree distributions, the GI network exhibits a significantly larger fraction of low-degree nodes. These nodes are predominantly found outside the largest opinion cluster, contributing to a reduced consensus size compared to the BA case.}
    \label{fig:degree_plots}
    \end{figure*}

Figure~\ref{fig:degree_plots} presents the scatter plots for both the BA and GI networks, giving the final opinion $x_i(t_{final})$ of all the agents, for a single realization, as a function of their degree $k$, ordered in decreasing order along the x-axis. Different vertical dotted lines correspond to different degree values (notice that for the Barabási–Albert network in Figure\ref{fig:BA} there are no nodes with degree less than $3$ by construction). The first striking observation is that the number of low-degree nodes present in the empirical GI network is much higher than in the BA topology of the same size, although both degree distributions are strongly heterogeneous, as indicated by the relative positions of the dotted vertical lines that delimit, on their right, nodes with $k\le K$,  for $K \in \{1,3,10,100\}$, respectively. In particular, the GI network has an excess of nodes with $k=1$ that are \textit{outsiders}, as they strongly populate the extreme opinions and do not belong to the largest opinion cluster, which is $x_S(t\rightarrow \infty) = 0.5$. Among these agents, those having an initial opinion difference from their single neighbor larger than 
$\varepsilon$  are excluded from the dynamics from the very first stages, thus diminishing the size of the final consensus cluster. As the initialization is uniform, the number of such nodes that can interact with their neighbor from the start increases linearly with $\varepsilon$. 

\begin{figure*}[htbp]
    \centering
    \includegraphics[width=0.9\textwidth]{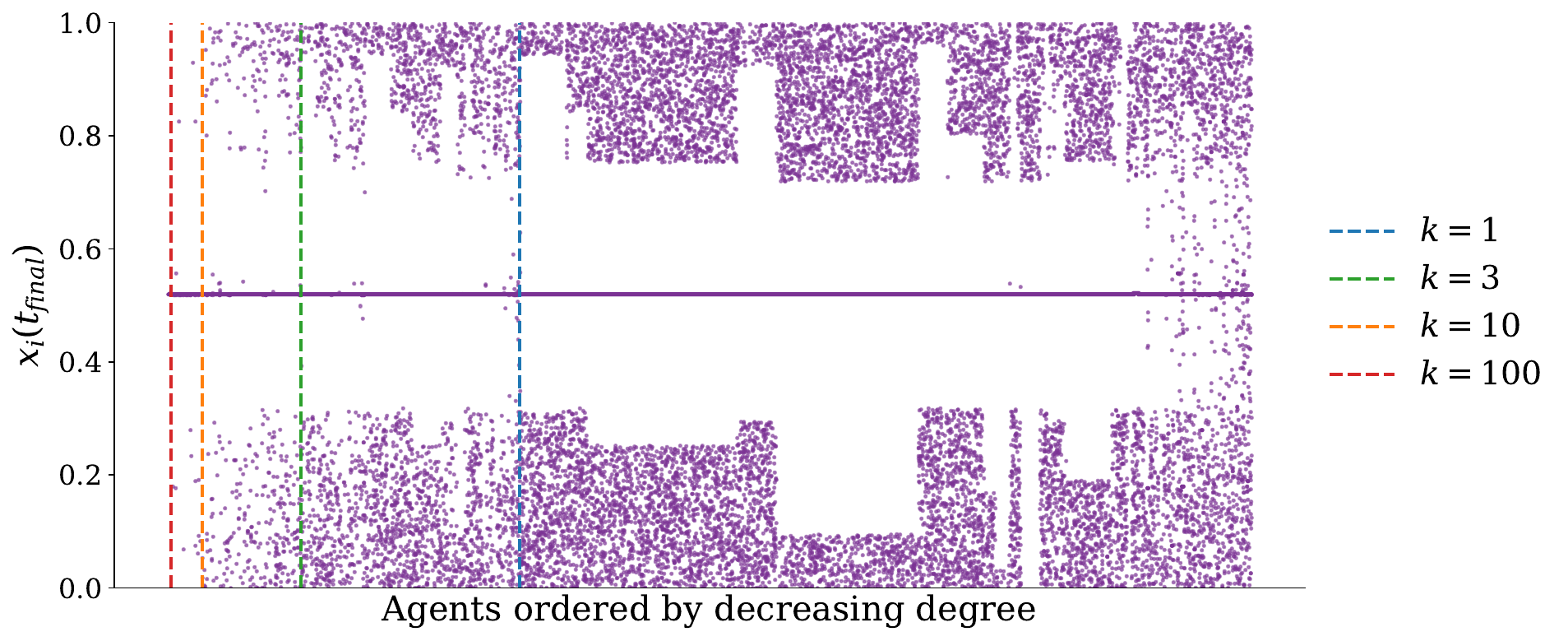} 
    \caption{Final opinions $x_i(t_{\mathrm{final}})$ as a function of node degree $k$ for a single realization of model $HK_1$ on the empirical GP network ($\varepsilon=0.2$, uniform initial conditions). Nodes are ordered by increasing degree along the horizontal axis, and vertical dotted lines indicate degree classes ($k \leq 1$, $1 < k \leq 3$, $3 < k \leq 10$, $10 < k \leq 100$, $k > 100$). A clear degree-dependent pattern emerges: highly connected nodes predominantly belong to the main opinion cluster, while low-degree nodes are more likely to remain outside it.
}
    \label{fig:GP scatterplot}
\end{figure*}

The same analysis can be done for the other empirical network used in this study, GP with similar conclusions as shown in Fig.~\ref{fig:GP scatterplot}.

We conclude, therefore, that for the studied empirical networks, the size of the largest cluster in the consensus region is determined by the high proportion of very low-degree nodes rather than by the community structure itself.

\section*{S6. $\Delta S$ behavior for GI and GP}

\begin{figure*}[htbp]
    \centering
    \includegraphics[width=\textwidth]{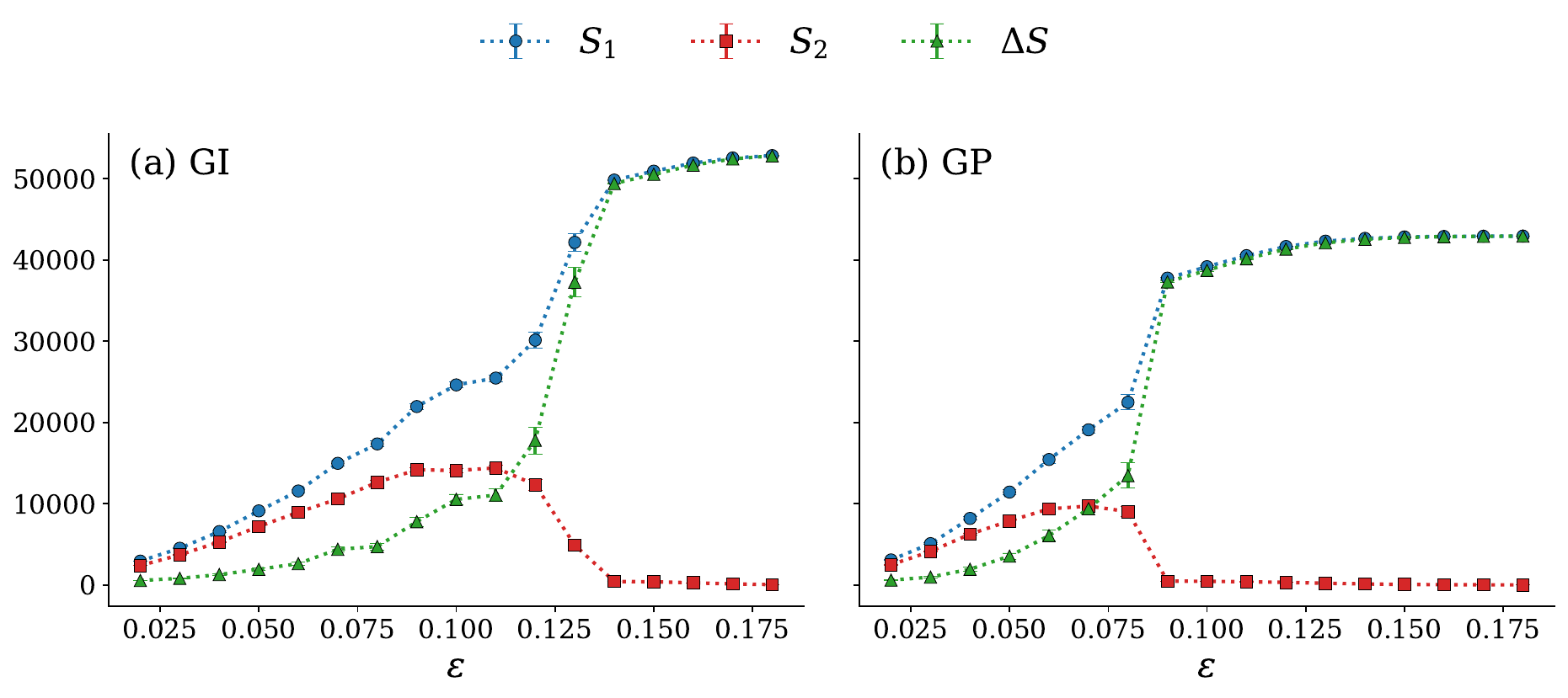} 
    \caption{
    Sizes of the two largest opinion clusters, $S_1$ and $S_2$, and their difference $\Delta S = S_1 - S_2$ as a function of the confidence bound $\varepsilon$ for model $HK_3$ ($\beta=3$, uniform initial opinions), averaged over $100$ realizations. Left: GI network. Right: GP network. 
    Unlike in artificial networks (ER and BA), $\Delta S$ does not decrease in the plateau region. Instead, it increases monotonically, indicating that $S_1$ and $S_2$ do not approach each other in size. This behavior suggests the absence of a merging process between the two largest clusters, with one cluster instead progressively dominating the system.
    }
    \label{fig:cluster_dynamics_empirical}
\end{figure*}

In this section, we present a detailed analysis of the origin of the \textit{plateau} in the order parameter of probabilistic models in stylized networks, which is absent in the empirical ones (see Fig.2 of the main text). While Fig.3 of the main text shows the sustained competition between the two largest opinion clusters in the polarization region, which gives rise to the \textit{polarization plateau} in the stylized networks, Fig.~\ref{fig:cluster_dynamics_empirical} shows the behavior of $S_1$, $S_2$, and $\Delta S$ as functions of $\varepsilon$ for the empirical networks GI and GP. 

Unlike in artificial networks, where the sustained competition between $S_1$ and $S_2$ generates the plateau region, here $\Delta S$ does not decrease in the plateau region. Instead, it continues to increase with $\varepsilon$, indicating that $S_1$ keeps growing faster than $S_2$. This suggests that, unlike in artificial networks, the two largest clusters do not undergo a merging process. Rather, one cluster progressively dominates the system without passing through a regime of comparable cluster sizes.

As a complement, we can quantify the variability in the difference in size, $\Delta S$, between the two largest clusters for all networks, using the entropy defined by Eq.(5) of the main text. The probability distribution of $\Delta S$ was computed by discretizing its values into bins of width 2000 agents using a common set of bin edges, spanning the interval $[0, 53002]$, for all network topologies. Using a fixed discretization ensures direct comparability of the entropy across datasets. Since bins outside the attainable range of a given network remain empty, they did not contribute to the Shannon entropy. Smaller bin widths were also tested, giving similar results. Therefore, $p_b$ indicates the probability associated with bin $b$. High entropy values indicate a broad diversity of possible $\Delta S$ outcomes, while low values correspond to more homogeneous realizations.

\begin{figure*}[htbp!]
  \centering
  \includegraphics[width=0.75\textwidth]{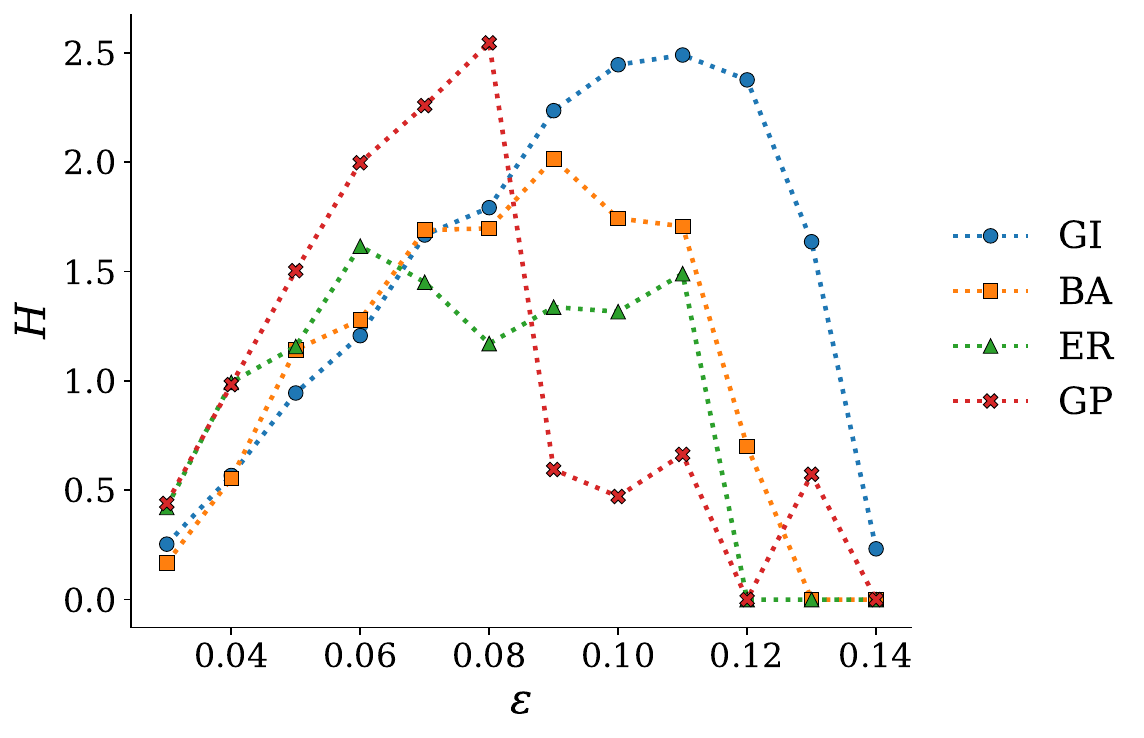}
  \caption{Shannon entropy $H$ of the distribution of $\Delta S = S_1 - S_2$ as a function of $\varepsilon$ for BA, ER, and the empirical networks GI and GP (model $HK_3$). Higher entropy values indicate greater variability of outcomes across realizations. The empirical networks display significantly larger entropy in the polarization region, explaining the absence of a well-defined plateau in the order parameter. Dotted lines are only a guideline for the eye.}
  \label{size_entropy}
\end{figure*}

\begin{figure*}[htbp!]
    \centering
    \begin{subfigure}[b]{0.45\textwidth}
        \centering
        \includegraphics[width=\textwidth]{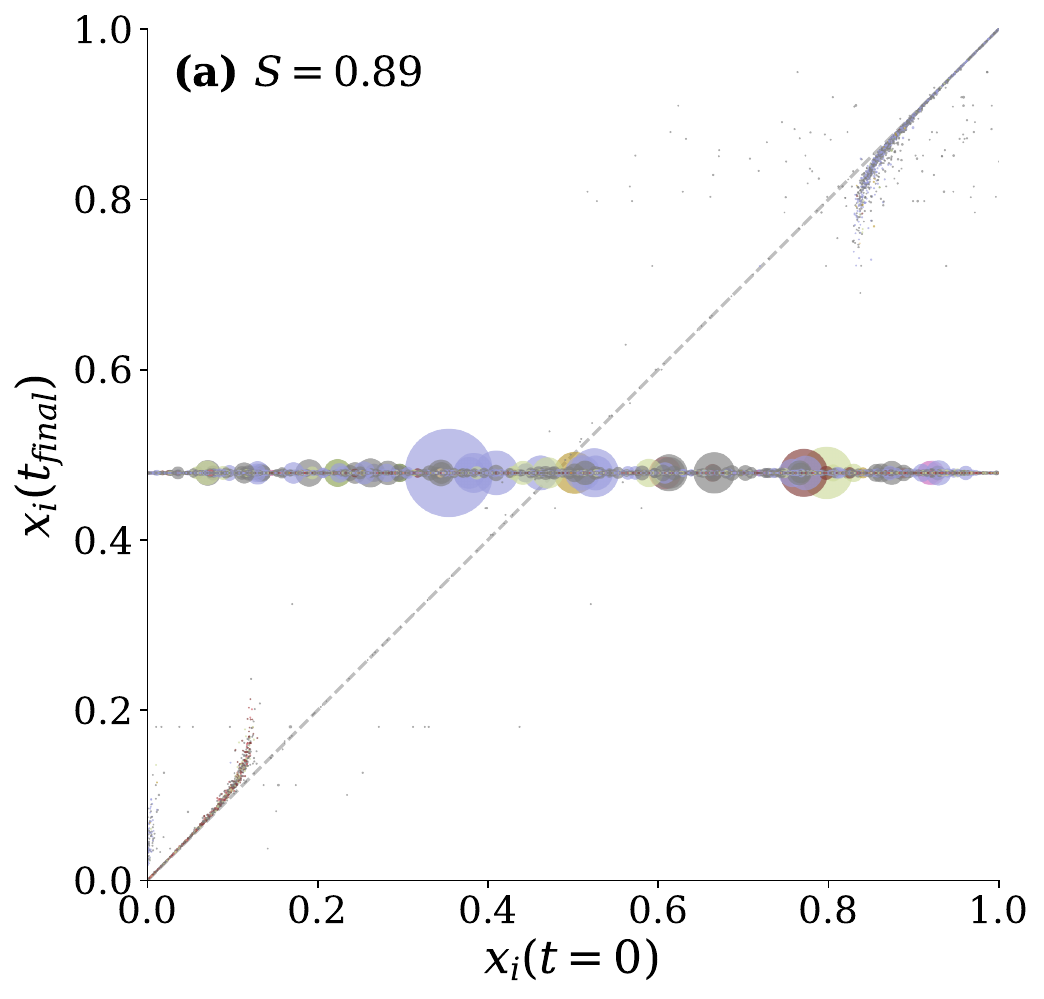}
        \caption{Consensus case $C$}
        \label{fig:consensus case}
    \end{subfigure}
    \hfill
    \begin{subfigure}[b]{0.45\textwidth}
        \centering
        \includegraphics[width=\textwidth]{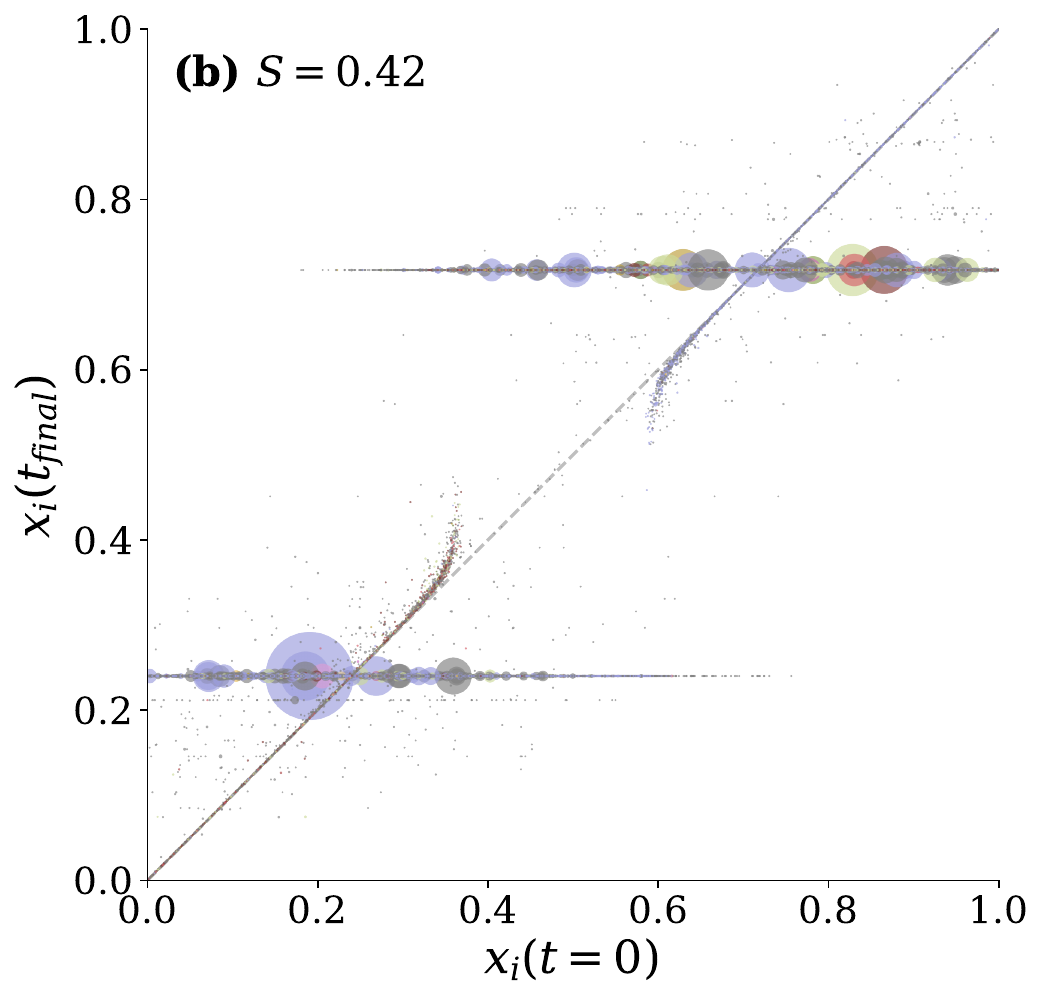}
        \caption{Non consensus case $NC$}
        \label{fig:non consensus case}
    \end{subfigure}
    \caption{Initial versus final opinions for two representative realizations of model $HK_3$ on the empirical GI network at $\varepsilon = 0.12$, with uniform opinions initialization. Left: a realization leading to consensus. Right: a realization remaining polarized. Each node is represented by a point whose size is proportional to its degree and whose color denotes community membership. In the consensus case, the highest-degree nodes are initially located near the center of the opinion space, facilitating convergence. In the non-consensus case, highly connected nodes are initially assigned more extreme opinions, hindering global consensus.}
    \label{fig:scatterplot opinions uniform}
\end{figure*}

As shown in Fig.~\ref{size_entropy}, in the region corresponding to the polarization plateau, the entropy of $\Delta S$ is significantly higher for the empirical networks than for the BA and ER networks, thus signaling high variability. Given that the topology of the empirical networks is fixed, the observed high entropy in the distribution of the size differences between the two largest opinion clusters in empirical networks can naturally be attributed to the  initialization. To test this hypothesis, we examine two individual realizations for $\varepsilon = 0.12$, one that reaches consensus and one that does not, in the GI empirical network. For this specific $\varepsilon$ value, the order parameter of $HK_3$ signals the existence of consensus; however, while many individual realizations end in a consensus state, several others end up quite far from it. Figure~\ref{fig:scatterplot opinions uniform} shows the scatter plot of final versus initial opinions for all nodes of two realizations representative of these situations. 

Fig.~\ref{fig:consensus case} shows that for the realization that reached consensus the most highly connected nodes are initially located near the center of the opinion space, favoring convergence by attracting many weakly connected neighbors toward moderate opinions. Conversely, in the non-consensus case, high-degree nodes are initially assigned more separated extreme opinions, which inhibits global consensus. These results indicate that in the empirical network, the particular initial opinions of the hubs can strongly influence the final outcome. This result suggests considering a different type of initialization that aligns with the real systems described by the empirical networks.

\section*{S7. General results for Gaussian initialization}

\begin{figure*}[t]
    \centering
    \begin{subfigure}[b]{0.49\textwidth}
        \centering
        \includegraphics[width=\textwidth]{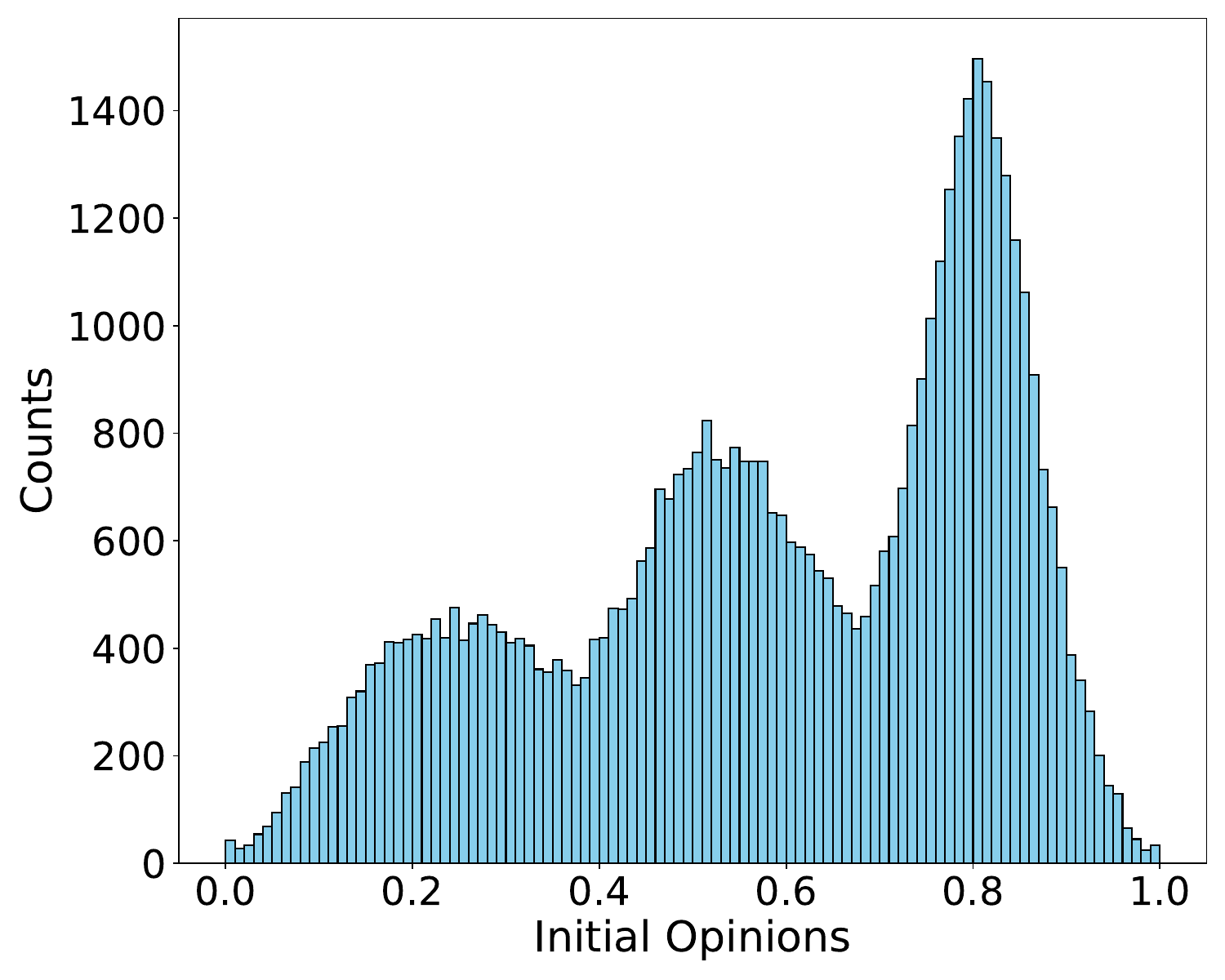} 
        \caption{GI}
        \label{fig:GI_gaussian}
    \end{subfigure}
    \hfill 
    \begin{subfigure}[b]{0.49\textwidth}
        \centering
        \includegraphics[width=\textwidth]{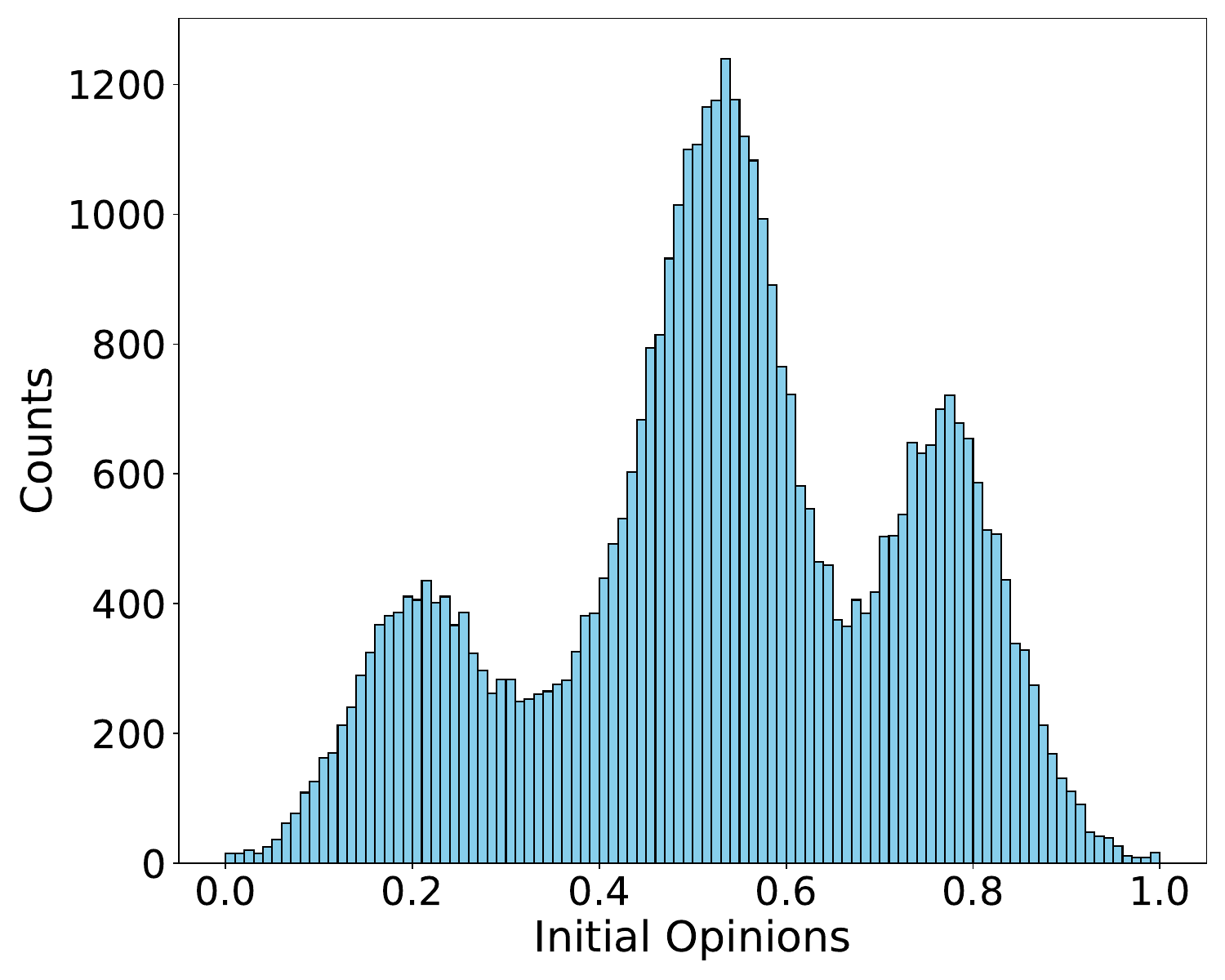} 
        \caption{GP}
        \label{fig:GP_gaussian}
    \end{subfigure}
    \caption{Example of Gaussian initial opinion distributions on the empirical networks GI (left) and GP (right). Opinions are assigned according to community membership, with values drawn from Gaussian distributions centered around community-specific means $\mu_k$ and standard deviation $\sigma=0.05$. 
}
    \label{fig:gaussian_example}
\end{figure*}

In the main text, we have extensively discussed Gaussian-distributed initial opinions on the empirical networks GI and GP. Figure \ref{fig:gaussian_example} illustrates an example for both networks, considering two random realizations. 

\FloatBarrier

\begin{figure*}[htbp]
    \centering
    \includegraphics[width=\textwidth]{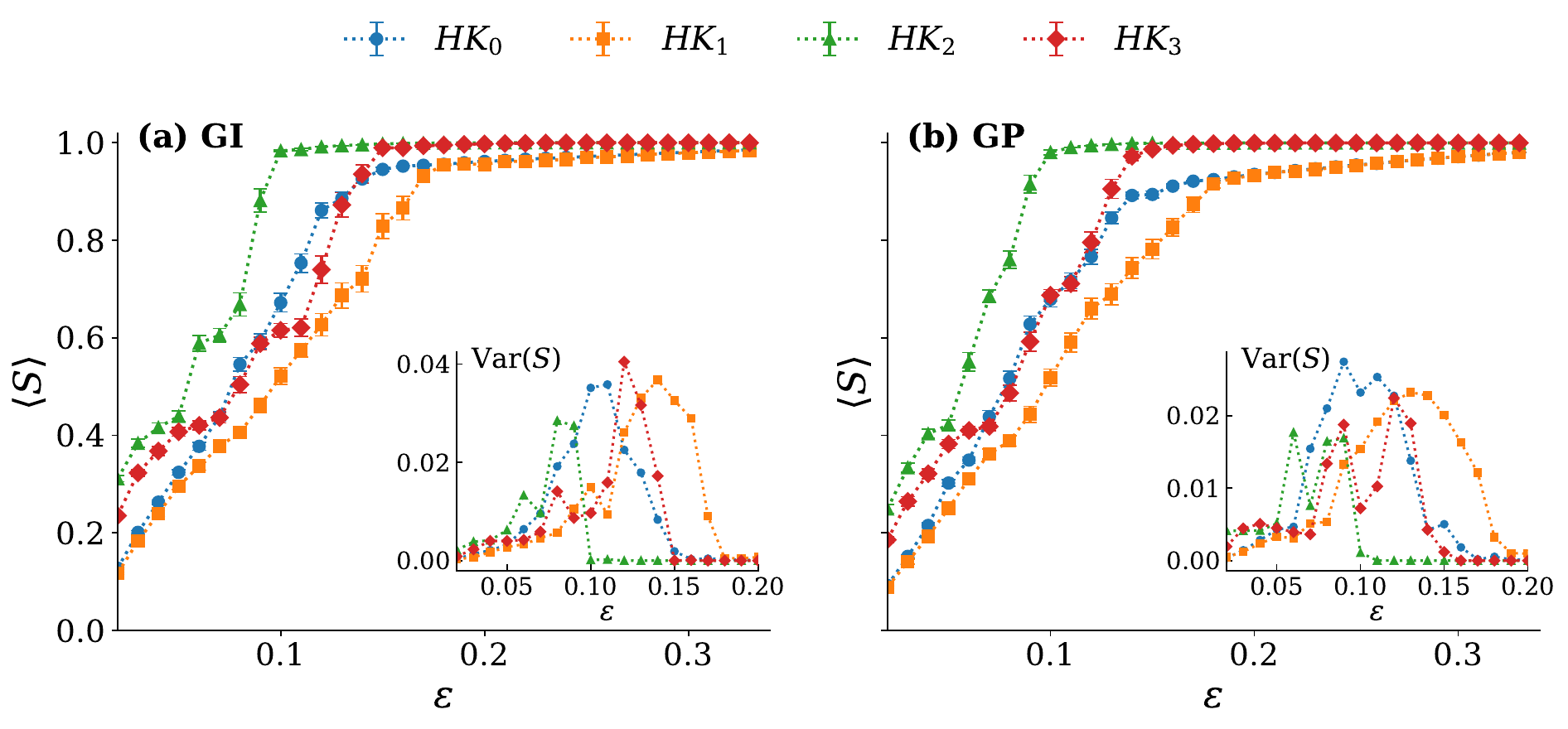} 
    \caption{Order parameter $<S>$ as a function of $\varepsilon$ for the four models ($HK_0$–$HK_3$) on the empirical GI (left) and GP (right) networks under Gaussian initial conditions. Insets show the variance used to estimate the critical threshold $\varepsilon_c$. The qualitative effects observed with uniform initialization are preserved: fuzzy neighbor selection lowers $\varepsilon_c$, promoting consensus, while weighted interactions increase $\varepsilon_c$, hindering consensus.
}
    \label{fig:S7_gaussian_consolidated}
\end{figure*}

Figure \ref{fig:S7_gaussian_consolidated} shows that the general findings presented in Fig.2 of the main text are robust to different initialization strategies and to different real networks. Specifically, the introduction of the fuzzy selection consistently lowers the critical threshold ($\varepsilon_c$) with respect to $HK_0$. Conversely, the enhancement of confirmation bias by means of weighted updates in the influence function systematically raises $\varepsilon_c$ compared with the corresponding case with uniform weights.

\begin{figure*}[t]
    \centering
    \includegraphics[width=\textwidth]{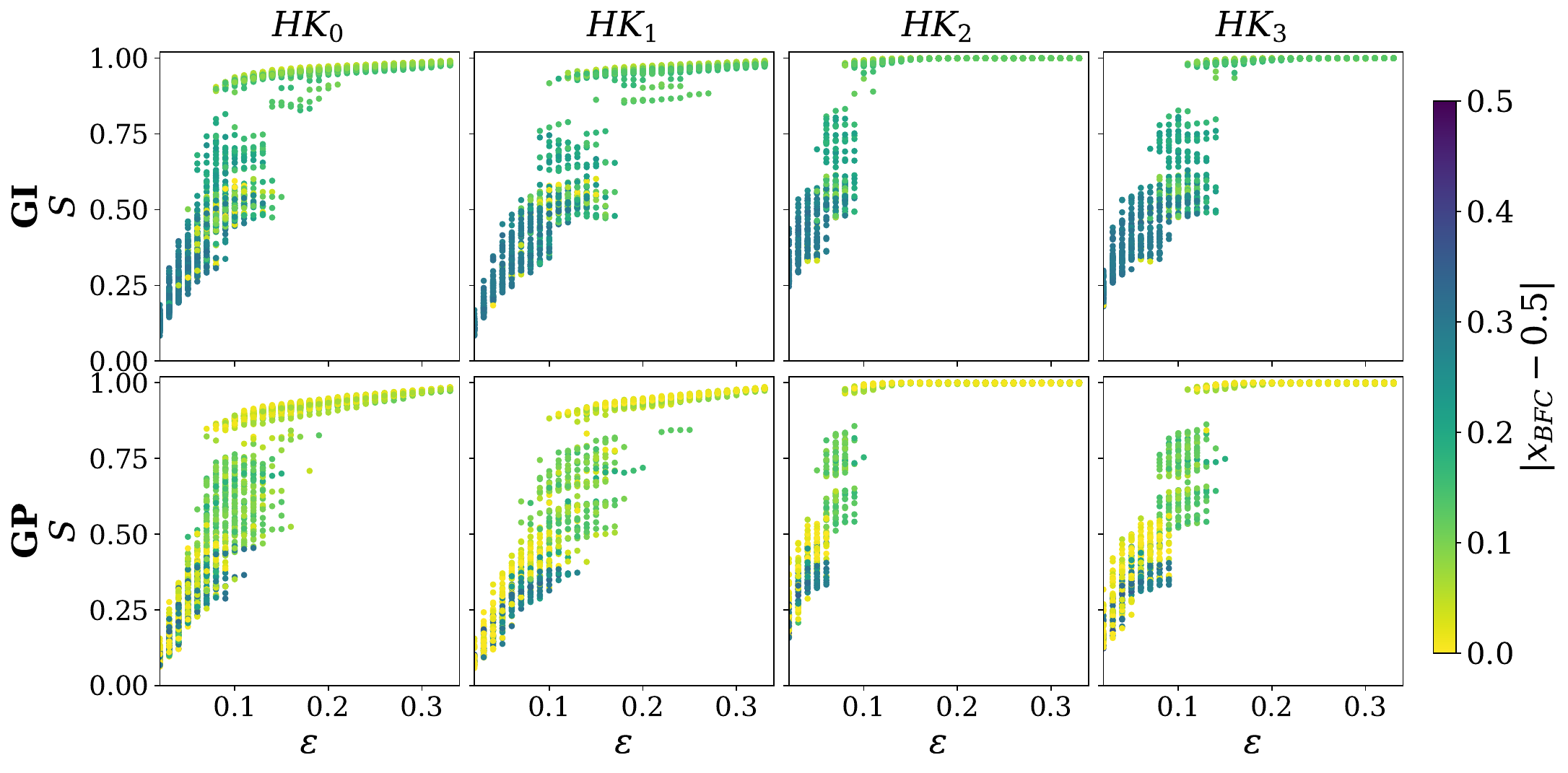} 
    \caption{Scatter plots of the normalized size $S$ of the largest opinion cluster for Gaussian initial conditions. Each column corresponds to a model and each row to an empirical network (GI, GP). Points represent independent realizations and are colored according to the final opinion of the largest cluster. Compared to the uniform case, in the Gaussian case, consensus can emerge at more extreme opinion values due to the community-based initialization.}
    \label{fig:scatterplots gaussian}
\end{figure*}

Beyond the average behavior, the Gaussian initialization also affects the distribution of outcomes across individual realizations. Figure~\ref{fig:scatterplots gaussian} reports the normalized size $S$ of the largest opinion cluster for each realization, together with its final opinion, for all four models and both empirical networks. Compared with the uniform initialization case, the final opinion of the largest cluster can reach more extreme values. This reflects the fact that the initial opinions are no longer homogeneously distributed over the opinion space, but are instead structured around community-specific means.

\FloatBarrier

\FloatBarrier

\begin{figure*}
    \centering
        \includegraphics[width=\textwidth]{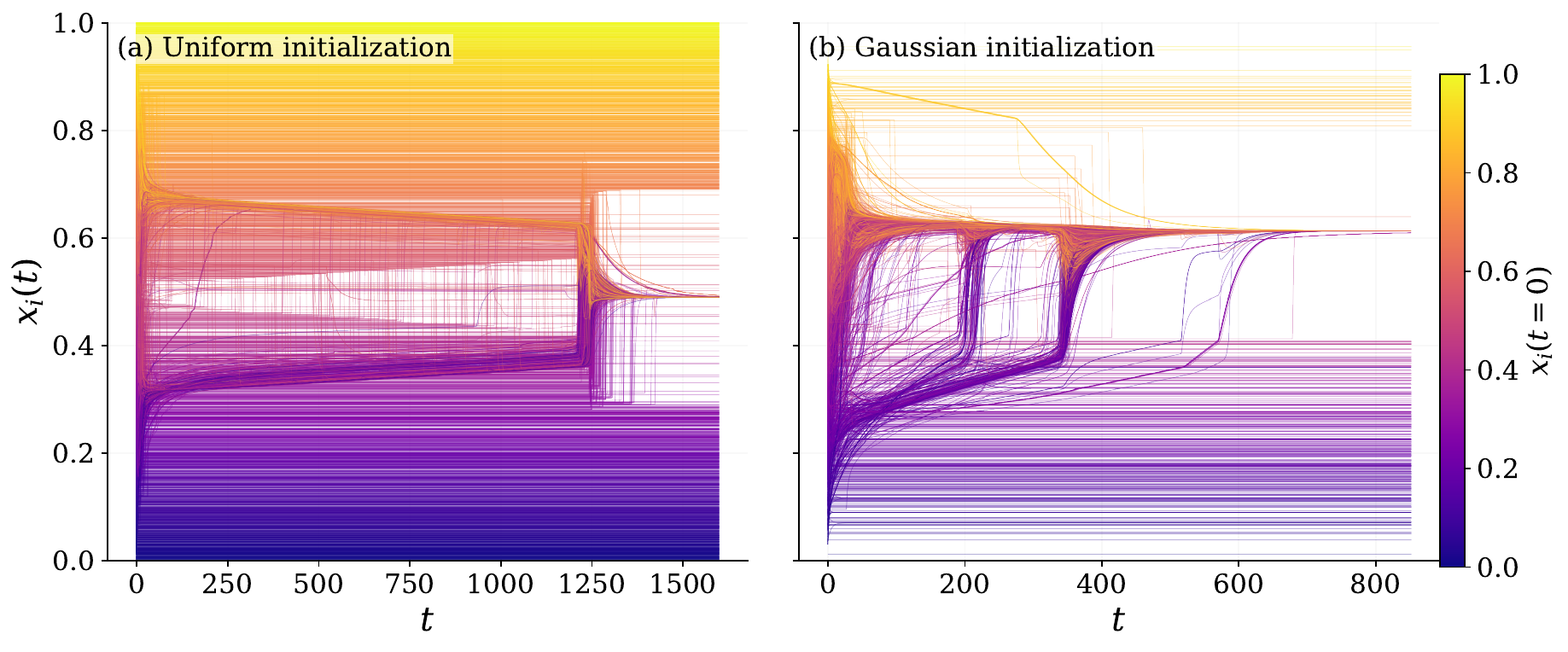}
    \caption{Opinion trajectories for a single realization of model $HK_1$ on the GI network at $\varepsilon=0.2$, comparing uniform (a) and Gaussian (b) initial conditions. For clarity, only $15\%$ of agents are shown. Trajectories are colored by initial opinion. Uniform initialization leads to convergence toward the center of the opinion space, whereas Gaussian initialization may result in a shifted consensus, highlighting the impact of structured initial conditions.}
    \label{fig:trajectories_gaussian_uniform}
\end{figure*}

This aspect can be better visualized using the opinion trajectories over time. Figure \ref{fig:trajectories_gaussian_uniform} illustrates the trajectories for a single realization of $HK_1$ on the empirical network GI for $\varepsilon=0.2$, where a clear consensus regime is observed. For better visibility, only 15\% of the total agents are shown.

The trajectories are color-coded according to the agent's initial opinion. In the uniform initialization case (Figure \ref{fig:trajectories_gaussian_uniform}(a)), opinions converge toward the center of the opinion interval ($\approx 0.5$), as expected. Conversely, in the Gaussian initialization case (Figure \ref{fig:trajectories_gaussian_uniform}(b)), the final consensus opinion is slightly shifted towards the $+1$ extreme. However, this specific result strongly depends on the GI network, the detected communities, and their initialization, and the outcomes differ when considering the GP network.

The initialization of opinions at the community level, and in particular that of highly connected nodes (hubs), plays a crucial role in the dynamics. While the main political communities have fixed mean opinions, the remaining ones are randomly initialized at each realization, leading to variability in the initial positions of hubs. This affects not only the final opinion of the largest cluster in the consensus regime, but also the outcome of the dynamics itself and, more specifically, the value of the order parameter for a single realization. For a fixed value of $\varepsilon$, different initializations can thus lead to significantly different values of $S$. This effect, already observed in the uniform initialization case, persists under Gaussian initialization.

To illustrate this, we fix the empirical network GI, the model $HK_3$, and $\varepsilon = 0.11$, and compare two representative realizations leading to different outcomes: one reaching consensus and one remaining polarized. While the means of the Gaussian distributions associated with the main political communities are fixed, the centers of all other communities are randomly reassigned at each realization, inducing variability in the initial positions of hubs.

\begin{figure*}[t]
    \centering
    \begin{subfigure}[b]{0.49\textwidth}
        \centering
        \includegraphics[width=\textwidth]{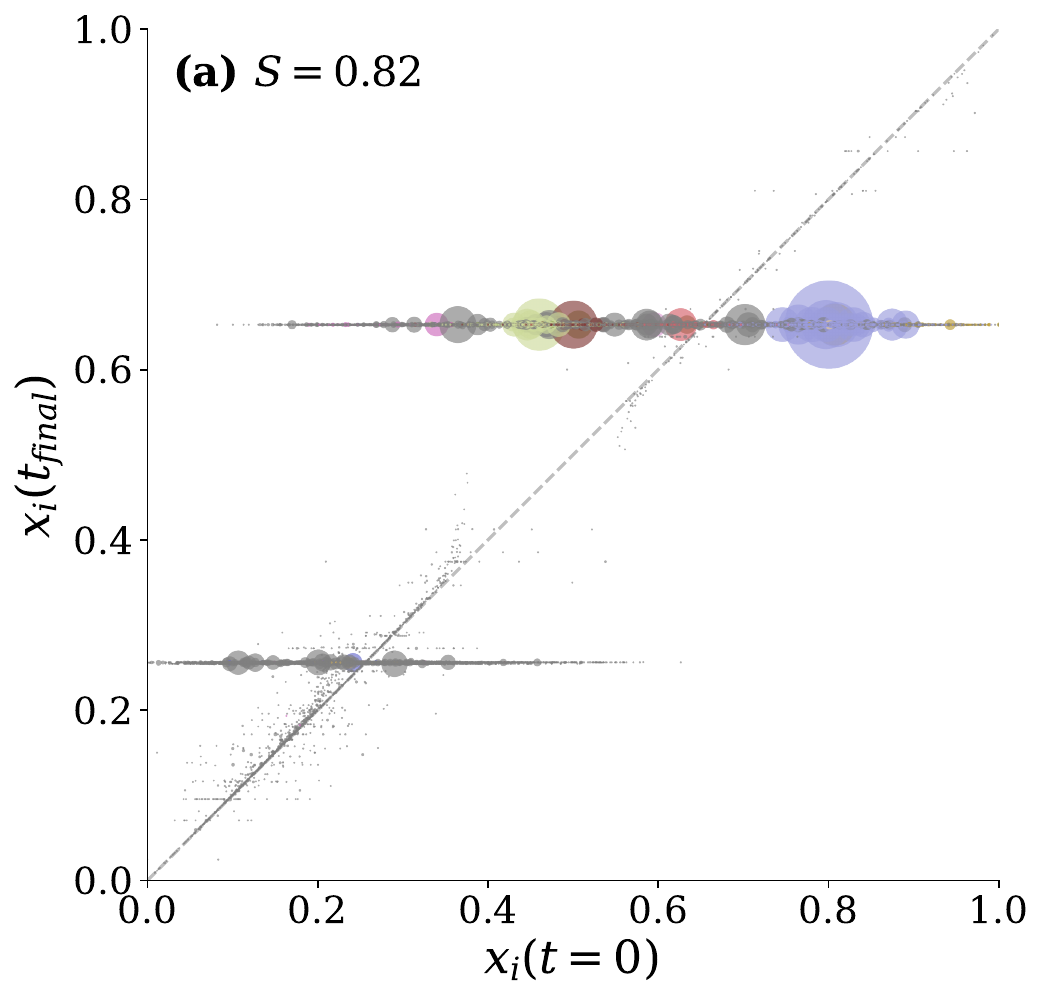} 
        \caption{Consensus case}
        \label{fig:consensus gaussian}
    \end{subfigure}
    \hfill 
    \begin{subfigure}[b]{0.49\textwidth}
        \centering
        \includegraphics[width=\textwidth]{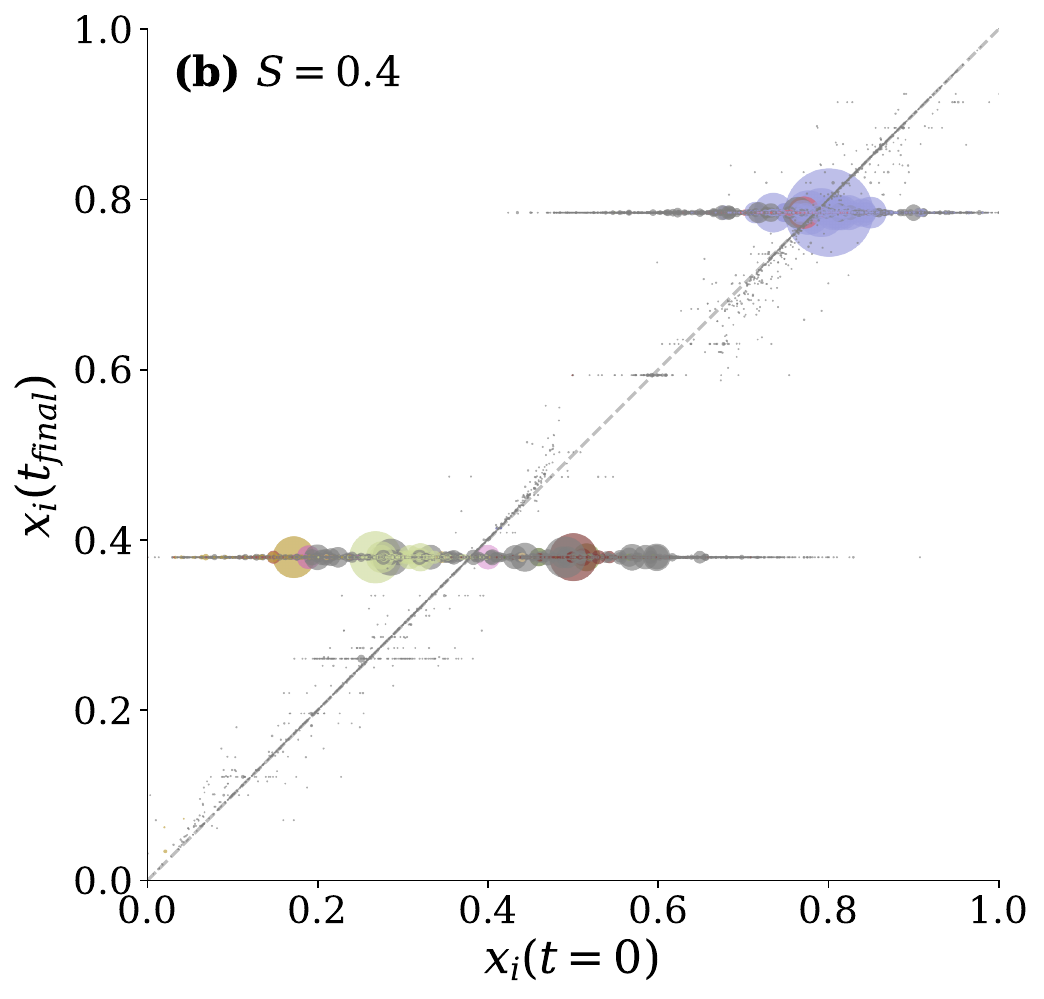} 
        \caption{Polarization case}
        \label{fig:non consensus gaussian}
    \end{subfigure}
    \caption{Initial versus final opinions for two representative realizations of model $HK_3$ on the empirical GI network at $\varepsilon = 0.11$, with Gaussian initial conditions. Each node is represented by a point whose size is proportional to its degree and whose color indicates community membership.}
    \label{fig:scatterplot opinions gaussian}
\end{figure*}

Figure~\ref{fig:scatterplot opinions gaussian} shows the scatter plots of initial versus final opinions for these two cases. Each node is represented by a point whose size is proportional to its degree and whose color indicates community membership.

\FloatBarrier

\FloatBarrier

The comparison highlights that different initial configurations of hubs can lead to qualitatively different outcomes. When hubs are initially aligned, they drive the system toward a common opinion, leading to consensus. Conversely, when their initial opinions are more dispersed, they can sustain competing clusters and prevent their merging, resulting in polarization.

\bibliography{sample}